\documentclass[11pt,a4paper]{article}
\usepackage[utf8x]{inputenc}
\usepackage[T1]{fontenc}
\usepackage{mathptmx} 

\usepackage[margin=25truemm]{geometry}
\usepackage[pdftex]{graphicx} 
\usepackage{calc} 
\usepackage{authblk}
\usepackage{bm}
\usepackage{enumitem} 
\usepackage{braket,mathtools,amsmath,amssymb,subcaption}
\usepackage{authblk}
\usepackage[all]{nowidow} 
\usepackage{url}

\usepackage{lipsum} 

\newcommand{\tl}[1]{\Tilde{#1}}

\usepackage{framed} 
\usepackage[framed]{ntheorem}
\newframedtheorem{frm-thm}{Theorem}

\def\hx{\hat{x}}
\def\hy{\hat{y}}
\def\hz{\hat{z}}
\def\tx{\tau^X}
\def\tz{\tau^Z}

\def\br{\mathbf{r}}

\def\bx{\mathbf{l}_x}
\def\by{\mathbf{l}_y}

\def\bp{\mathbf{p}}

\def\ex{\frac{\mathbf{e}_x}{2}}
\def\ey{\frac{\mathbf{e}_y}{2}}

\def\ea{\frac{\mathbf{e}_a}{2}}
\def\eb{\frac{\mathbf{e}_b}{2}}
\def\exx{\mathbf{e}_x}
\def\eyy{\mathbf{e}_y}
\def\bc{\mathbf{c}}
\def\ezz{\mathbf{e}_z}

\usepackage{hyperref}
\hypersetup{colorlinks,linkcolor=magenta,citecolor=cyan,urlcolor=cyan}
\begin{document} 
\title{
Noninvertible 
operators in 
one, two, and three dimensions
via gauging spatially modulated symmetry}

\author{Hiromi Ebisu$^1$}
\author{Bo Han$^2$}
\affil{$^1$Yukawa Institute for Theoretical Physics, Kyoto University, Kyoto 606-8502, Japan}
\affil{$^2$Department of Condensed Matter Physics, Weizmann Institute of Science, Rehovot 7610001, Israel}
                            
\maketitle
\thispagestyle{empty}

\begin{abstract}
Spatially modulated symmetries have emerged since the discovery of fractons, which characterize unconventional topological phases with mobility-constrained quasiparticle excitations. On the other hand, noninvertible symmetry operators have attracted substantial attention in communities of high energy and condensed matter physics due to their deep insight into quantum anomalies and exotic phases of matter. However, the connection between these exotic and noninvertible symmetries has not been
fully explored. In this paper, we construct
concrete lattice
models with noninvertible symmetry operators via gauging spatially modulated symmetries and investigate their exotic fusion rules. Specifically, we construct spin models with {\it subsystem symmetries} or {\it dipole symmetries} on one, two, and three-dimensional lattices. Gauging subsystem symmetries leads to noninvertible symmetry operators whose fusion rules involve $0$-form subsystem charges in two dimensions and higher-form operators that correspond to ``lineon'' excitations~(excitations which are mobile along one-dimensional line) in three dimensions. Gauging dipole symmetries leads to noninvertible symmetry operators with {\it dipole algebras} that describe a hierarchical structure between global and dipole charges. Notably, the hierarchical structure of the dual dipole charges is inverted compared with the original ones. Our work provides a unified and systematic analytical framework for constructing exotic symmetry operators by gauging relevant symmetries.
\end{abstract}

\newpage
\pagenumbering{arabic}

\tableofcontents
\section{Introduction}
Symmetry has been an important guiding principle in physics, allowing us to gain deeper insight of problems in various contexts---it imposes powerful constraints on a theory, from which one can predict many properties without explicit calculations. Recently, the concept of symmetries has been broadened, initiated by works~\cite{doi:10.1073/pnas.0803726105,gaiotto2015generalized} on generalized symmetries associated with extended objects. 
With the advent of fractonic topological phases~\cite{chamon,Haah2011,Vijay}, new types of symmetries, namely, {\it spatially modulated symmetries} (a.k.a. \textit{exotic symmetries}) emerge, where unconventional topological orders admit quasiparticle excitations with mobility constraints. 
In bosonic theories, the system has U(1) spatially modulated symmetry if the Hamiltonian is invariant under the transformation on the boson field $b_\br$ as $b_\br \to e^{i \theta(\br)} b_\br$, where $\theta(\br)$ depends on the lattice site $\br$. For instance, choosing $\theta (\br) = a$ leads to the regular global U(1) symmetry; choosing $\theta(\br) = \boldsymbol{b} \cdot \br$ leads to the dipole U(1) symmetry. Here $a$ and $\boldsymbol{b}$ are constants. The corresponding interactions with the global and \textit{dipole} U(1) symmetries are  $b^\dagger_{\br} b_{\br'}$ and $b^\dagger_\br b_{\br+\hat{x}} b^\dagger_{\br+\hat{x} + \hat{y}} b_{\br + \hat{y}}$, respectively. Similarly, taking $\theta(\br)$ as a step function leads to a \textit{subsystem symmetry}.\footnote{There is an important distinction between dipole and subsystem symmetries. In particular, space transformation act differently on unitary operators of these two symmetries.  We denote $U(\Sigma)$ as a unitary operator that implements the unitary transformation dictated by $U$ on submanifold $\Sigma$. In the case of a subsystem symmetry, 
the unitary operator transforms as 
\begin{equation}
    S U(\Sigma)S^\dagger=U(\Sigma_S), 
\end{equation}
where $\Sigma_S$ represents another submaifold related by $\Sigma$ via spatial transformation $S$. On the other hand, in the case of a dipole symmetry, the unitary operator transforms as 
\begin{equation}
    S U(\Sigma)S^\dagger =U_S(\Sigma_S)\label{ppq}
\end{equation}
In contrast to the transformation under a subsystem symmetry, $U_S$ is a different unitary transformation on the submanifold $\Sigma_S$. If one defines 
the modulated symmetry as a symmetry for which there is a group action of space symmetries on the internal symmetries such as the relation~\eqref{ppq}, dipole symmetries are modulated symmetries whereas subsystem symmetries are {\it not}. However, we regard both subsystem and dipole symmetries as modulated symmetries based on the fact that both have inhomogeneous symmetry transformations. An example for this inhomogeneous symmetry transformation is $\theta(\br)$ in the case of the bosonic model explained above in the main text. This perspective is also adopted in the literature. See Ref.~\cite{Sala2022sym} for example. We thank the anonymous referee for pointing out this distinction.
}
\par
In this work, we focus on two examples of the spatially modulated symmetries --
subsystem symmetry~\cite{Ring2002,plqt_ising_2004,Vijay,seiberg2021exotic}, and multipole symmetries~\cite{griffin2015scalar,Pretko:2018jbi,PhysRevX.9.031035,Pretko_dis}.
In the former symmetry, 
a symmetry defect is topological in a particular direction, rather than entire space.
Topological phases with subsystem symmetries possess a subextensive number of conserved charges, and the ground-state degeneracy (GSD) becomes subextensive in fracton topological phases with subsystem symmetry, such as the X-cube model~\cite{Vijay,foliated1,shirley2019fractional}. 
Subsystem symmetries also lead to particle excitations with mobility constraints, such as ``lineons'' or ``planons'' (excitations which are mobile on a line or a plane, respectively).
Multipole symmetries are associated with conservation of multipoles, such as dipole symmetries.
These symmetries put mobility constrains on excitations, giving rise to unusual GSD dependence on the system size in gapped phases, which was not observed for subsystem symmetries~(see e.g., \cite{PhysRevB.106.045112,PhysRevB.106.045145,oh2022rank}). Furthermore, 
multipole symmetries yield new insights in quantum field theories, allowing us to explore
various research directions, such as multipolar fractional excitations~\cite{2023foliated,2024multipole}, 
 Hilbert space fragmentations in dynamically constrained systems~\cite{PhysRevX.10.011047,Moudgalya_2022}, and hydrodynamics~\cite{PhysRevResearch.2.033124,PhysRevResearch.3.043186}.\par
Along a different stream of research, \textit{noninvertible symmetries} (a.k.a. categorical symmetries) 
have attracted a plethora of attentions~(see e.g.,~\cite{PhysRev.60.252,Frohlich:2006ch,Feiguin2007,Aasen_2016,Chang:2018iay,thorngren2019fusion,inamura2021topological}). One familiar example of such a symmetry is the Kramers-Wannier duality defect in the Ising model~\cite{PhysRev.60.252}. 
Unified framework to deal with such symmetries has been developed by making use of fusion category theories~\cite{TAMBARA1998692,etingof2005fusion}. 
Although theoretical frameworks of the noninvertible symmetries in one dimension are well-established, elucidation of this type of symmetries in higher dimensions is still an active area of research~\cite{douglas2018fusion,Gaiotto:2019xmp,Koide:2021zxj,Choi:2021kmx,Kaidi:2021xfk,Bhardwaj:2022yxj,Delcamp:2023kew,Bhardwaj:2022kot}. \par

Among various kinds of symmetries, it is desirable to establish a theoretical framework with broader scope, incorporating these symmetries in a unified fashion. 
On the other hand, there are growing interests in explicit constructions of noninvertible symmetry operators in lattice models.
Spurred by these facts, in this paper, we explore intertwining of the two types of symmetries, i.e., noninvertible symmetries and spatially modulated symmetries by constructing exactly solvable lattice models of noninvertible operators by gauging spatially modulated symmetries in one, two, and three dimensions. Our construction is motivated by recent progress of noninvertible operators in two dimensional systems comprised of double copies of spin models~\cite{Choi:2024rjm}. 
We find that noninvertible operators obtained by gauging subsystem or dipole symmetries exhibit rich structures in fusion rules. 
For subsystem symmetries, we show that fusion rules of the operators are characterized by subsystem charges and higher form operators corresponding to the ``lineon'' excitations which are studied in the X-cube model~\cite{Vijay}, the simplest model of the fracton topological phase. Such fusion rules are demonstrated in~\eqref{33}~\eqref{58}. 
For dipole symmetries, the operators exhibit a new type of dipole algebras, which describes how global and dipole charges are related via translational operators. We show that the hierarchical structure of such algebra plays a crucial role to understand fusion rules of the operators. Such fusion rules are summarized in~\eqref{77},~\eqref{fusiondip},~\eqref{fusiondip2},~\eqref{fusiondip3}, and~\eqref{fusiondip4}.
\par 

Note that explorations of noninvertible 
operators via gauging
spatially modulated symmetries have recently started (see e.g.,~\cite{Cao:2023rrb,Cao:2023doz,ParayilMana:2024txy,Spieler:2024fby,Cao:2024qjj,Pace:2024tgk,Pace:2024acq,Seo:2024its}). Yet complete understanding of
these operators, in particular operators in two and higher dimensions, remains elusive. 
Our work fills this gap and complies with diverse interests in various symmetries.\par
The rest of this work is organized as follows. 
In Sec.~\ref{section2}, we go over how noninvertible operator is obtained by the doubled copies of the spin model, the detail of which is crucial in the subsequent discussion. 
In Sec.~\ref{section3}, we construct noninvertible operator in two and three dimensions obtained by gauging subsystem symmetries. In Sec.~\ref{section4}, we turn our attention to the case of the dipole symmetries. After emphasizing the importance of the dipole algebra, we argue that fusion rules of noninvertible operators via gauging dipole symmetries involves such algebraic structure. 
In Sec.~\ref{section5}, we conclude our work with commenting on future perspectives. Technical issues are relegated into Appendix. 




\section{Conventional Ising chain}\label{section2}
In this section, we briefly recap the argument presented in~\cite{Choi:2024rjm} to obtain systematically noninvertible operators. 
Specifically, we review noninvertible operators in double copies of the Ising chains, which also serves as the basic approach for our systematic constructions of noninvertible operators of spatially modulated symmetries in Sec.~\ref{section3} and \ref{section4}.\par
We start our discussion from double Ising chains
\begin{eqnarray}
    H=-J\sum_jZ_jZ_{j+1}-h\sum_jX_j-\tl{J}\sum_j\tl{Z}_j\tl{Z}_{j+1}-\tl{h}\sum_j\tl{X}_j, 
    \label{ising}
\end{eqnarray}
where $X_j (\tilde{X}_j)$ and $Z_j (\tilde{Z}_j)$ are the two independent copies of Pauli operators on site-$j$. 
We take the periodic boundary condition $O_{j+L} = O_{j}$, where~$L$ is the system size and $O$ can be any Pauli operator. 
The Hamiltonian (\ref{ising}) has the following two global symmetries for generic couplings $J(\tilde{J}), h (\tilde{h})$
\begin{eqnarray}
    Q=\prod_{j=1}^LX_j,\quad   \tl{Q}=\prod_{j=1}^L\tl{X}_j.
\end{eqnarray}
At the special point where $h=\tl{h}$ and $J=\tl{J}$, the system has an additional $\mathbb{Z}_2$ symmetry, 
which exchanges~$O_j$ and $\tl{O}_j$. 
Specifically, we introduce the following swap operator:
\begin{eqnarray}
    S_j\vcentcolon=\frac{1}{2}(I+X_j\tl{X}_j+Z_j\tl{Z}_j-Z_j\tl{Z}_jX_j\tl{X}_j)\label{sj}
\end{eqnarray}
whose action on a local spin reads
\begin{eqnarray}
    S_jX_j=\tl{X}_j S_j,\quad  S_j\tl{X}_j=X_j S_j,\quad  S_jZ_j=\tl{Z}_j S_j,\quad  S_j\tl{Z}_j=Z_j S_j.
\end{eqnarray}
From now on, we assume $h=\tilde{h}$ and $J = \tilde{J}$, which leads to the $\mathbb{Z}_2^{swap}$ symmetry generated by $S = \prod^L_{j=1} S_j$. 
As a result, the model respects the $D_8$ symmetry, whose algebra is represented as 
\begin{eqnarray}
[H,Q]=[H,\tl{Q}]=[H,S]=0,\quad SQ=\tl{Q}S,\quad S\tl{Q}=QS,\quad S^2=Q^2=\tl{Q}^2=1,\quad Q\tl{Q}=\tl{Q}Q.~\label{qq}    
\end{eqnarray}
The discussion in what follows remains valid as long as the model has this symmetry. 
One could add admissible terms to respect the symmetry~\eqref{qq} to the Hamiltonian~\eqref{ising}.
Examples of such terms are
\begin{eqnarray*}
   -g_1\sum_jZ_jZ_{j+1}\tl{Z}_j\tl{Z}_{j+1}-g_2\sum_jX_j\tl{X}_j,
\end{eqnarray*}
the first of which was considered in the Ashkin-Teller model~\cite{Ashkin_Teller}.
\par
To construct noninvertible operators, we gauge one of the global symmetries, say, $\tl{Q}$. To do this, we introduce extended a Hilbert space, corresponding to gauge field, described by $\tl{\tau}^X_{j+1/2}$, $\tl{\tau}^Z_{j+1/2}$ located on each link. The Gauss law reads\footnote{Alternatively, one performs a unitary transformation (see e.g.,\cite{Cobanera:2009as}) that simplifies the Gauss law term to a local single $\mathbb{Z}_2$ operator, say, $\zeta^X_j$ and set $\zeta^X_j=1$. 
}
\begin{eqnarray}
    \tl{\tau}^X_{j-1/2}\tl{X}_j\tl{\tau}^X_{j+1/2}=1. \label{eqn:gauss}
\end{eqnarray}
Further, the spins are minimally coupled to the gauge field as
\begin{eqnarray}
    \tl{Z}_j\tl{Z}_{j+1}\to \tl{Z}_j\tl{\tau}^Z_{j+1/2}\tl{Z}_{j+1}.
\end{eqnarray}
Defining new gauge invariant variables as 
\begin{eqnarray}
    \tau^X_{j+1/2}\vcentcolon=\tl{\tau}^X_{j+1/2},\quad \tau^Z_{j+1/2}\vcentcolon=\tl{Z}_j\tl{\tau}^Z_{j+1/2}\tl{Z}_{j+1}\label{variables}
\end{eqnarray}
we have the following mapping via gauging the global symmetry: 
\begin{eqnarray}
    \tl{X}_{j}\Rightarrow G^X_{j},\quad \tl{Z}_{j}\tl{Z}_{j+1}\Rightarrow \tau^Z_{j+1/2}\label{mp}
\end{eqnarray}
where 
\begin{eqnarray*}
    G^X_{j}\vcentcolon=\tau^X_{j-1/2}\tau^X_{j+1/2}.
\end{eqnarray*}
Here, the arrow ``$\Rightarrow$''represents mapping between the operators via gauging 
\footnote{We explain more about ``$\Rightarrow$'' as it will be used throughout this paper. For operators $\hat{A}$ and $\hat{B}$, we use $\hat{A} \Rightarrow \hat{B}$ to imply that after gauging, $\hat{A}$ should be replaced by $\hat{B}$ in the gauged theory. Specifically, in Eq.~(\ref{mp}), the first relation results from solving $\tilde{X}_j$ in the Gauss' law Eq.~(\ref{eqn:gauss}), after which $\tilde{X}_j$ will be replaced with $G^X_j$ in the following discussions. The second relation in Eq.~(\ref{mp}), on the other hand, does not result from solving any constraint relation. Instead, it is a redefinition of the gauge field $\tilde{\tau}^Z_{j+1/2} \to \tau^Z_{j+1/2}$ through~\eqref{variables}, after which all the commutation relations are preserved.}. 
The gauged Hamiltonian reads
\begin{eqnarray}
    \widehat{H}=-J\sum_jZ_jZ_{j+1}-h\sum_jX_j-{J}\sum_j\tau^Z_{j+1/2}-{h}\sum_jG^X_j.
\end{eqnarray}
This Hamiltonian admits the following dual symmetry:~\footnote{It is sometimes referred to as \textit{quantum symmetry}~\cite{Vafa1989quantumsym}.}
\begin{eqnarray}
    \eta\vcentcolon=\prod_{j=0}^{L-1}\tau^Z_{j+1/2}.
\end{eqnarray}
\par
Before gauging the global symmetry, the operator $S$ sends a local spin $X_j$ and $Z_j$ to $\tl{X}_j$ and~$\tl{Z}_j$, respectively. These two are further mapped to~$G^X_{j}$ and $\tau^Z_{j+1/2}$ after gauging, according to~\eqref{mp}. Hence, one naively expects that the operator $S$ plays a role of the operators; operator $X_j$ is mapped to~$G^X_{j}$ and $Z_jZ_{j+1}$ to~$\tau^Z_{j+1/2}$, which can be seen in the typical duality transformation in the Ising model. 
However, one faces an issue; from the form of $S_j$~\eqref{sj}, a local operator $\tl{Z}_j$ in $S_j$ does not commute with the global symmetry~$\tl{Q}$. Hence, after gauging, it should become non-local operator, involving the gauge fields~$\tau^Z_{j+1/2}$. Based on the facts that the swap operator $S$ should behave as the operator, and that a local operator $\tl{Z}_j$ inside the swap operator becomes nonlocal after gauging, we replace~$\tl{Z}_j$  with the Wilson operator $W_j$ consisting of string of the gauge fields: 
\begin{eqnarray}
   W_j=\prod_{i=0}^{j-1}\tau^Z_{i+1/2}.\label{10}
\end{eqnarray}
Accordingly, the operator $S_j$ after gauging becomes
\begin{eqnarray}
    S_j=\frac{1}{2}\left[(1+G^X_jX_j)+Z_j\left(\prod_{i=0}^{j-1}\tau^Z_{i+1/2}\right)(1-G^X_jX_j)\right].\label{eqn:gaugedS}
\end{eqnarray}
Note that $S_j~(1\leq j\leq L-1)$ is invertible but $S_L$ is noninvertible. 
Indeed, 
$S_j^2=1~(1\leq j\leq L-1)$ whereas $S_L^2=S_L$. 
Also, $\{S_j|1\leq j\leq L-1\}$ commutes with one another whereas $S_L$ does not with $S_{j}\;(1\leq j\leq L-1)$.
Define product of the swap operator as $S=\prod_{j=1}^LS_j$, 
after some algebra one has 
\begin{align}
    &SX_j=\begin{cases}
        G^X_jS\quad(1\leq j\leq L-1)\\
        G^X_LSQ\quad(j=L)
    \end{cases}&,&\quad&&
    SG^X_j=X_jS\nonumber\\
    &SZ_jZ_{j+1}=\begin{cases}
        \tau^Z_{j+1/2}S\quad(1\leq j\leq L-1)\\
        \tau^Z_{1/2}S\eta\quad(j=L)
    \end{cases}&,&\quad&&
        S\tau^Z_{j+1/2}=\begin{cases}
        Z_jZ_{j+1}S\quad(1\leq j\leq L-1)\\
        Z_LZ_1S\eta\quad(j=L).
    \end{cases}\label{cs1}
\end{align}
%
Each relation in~\eqref{cs1} can be verified by a simple algebra. For instance, given $S X_i = \tau^x_{i-1/2} \tau^x_{i+1/2}$ for $i = 1,\dots, L-1$, we have
\begin{align}
    \tau^x_{L-1/2} \tau^x_{1/2} S &= (\tau^x_{L-1/2} \tau^x_{L-3/2}) \cdot (\tau^x_{L-3/2} \tau^x_{L-5/2}) \cdots (\tau^x_{3/2} \tau^x_{1/2}) S \nonumber \\
    &= S X_{L-1} X_{L-2} \cdots X_1 = S X_L Q,
\end{align}
which is essentially
\begin{align}
    S X_L &= G^X_L S Q.
\end{align}

The relations in~\eqref{cs1} indicates that away from the boundary, $S$ indeed plays a role of the mapping corresponding to gauging~\eqref{mp}; yet the action of $S$ on spin at the boundary $(j=L)$ leads to non-local charges. To remedy this issue, one multiplies global charges with $S$ by introducing 
\begin{eqnarray}
    D\vcentcolon=\frac{1}{2}S(1+Q)(1+\eta),
\end{eqnarray}
which acts as the desired swap operator:
\begin{eqnarray}
    DX_j=G^X_jD,\quad DG^X_j=X_jD,\quad DZ_jZ_{j+1}=\tau^Z_{j+1/2}D,\quad D\tau^Z_{j+1/2}=Z_{j}Z_{j+1}D
\end{eqnarray}
Furthermore, it is noninvertible, as is evident from the fusion rules of the operators
\begin{eqnarray}
    D\times D=(1+Q)(1+\eta),\quad QD=DQ=D,\quad \eta D=D\eta=D.\label{rep8}
\end{eqnarray}
The fusion rules~\eqref{rep8} exhibits the noninvertible 
Rep$(D_8)$ symmetry.\par
In the subsequent sections, we follow the approach discussed here for Ising models to construct noninvertible operators via gauging spatially modulated symmetries in lattice models. 
The strategy to establish such operators is proceeded in the following three steps: 
\begin{enumerate}
  \item Introduce double copies of spin models, each of which has a global (spatially modulated) symmetry and additional $\mathbb{Z}_2$ symmetry exchanging the two spin degrees of freedom. One could add other interacting terms to the model as long as the Hamiltonian respects these symmetries. [Shown in Eq.~(\ref{ising}) to (\ref{qq}).]
  \item Gauge one set of the global symmetries for one spin degree of freedom, and express the spins in the gauged copy by the gauge fields in the swap operators; in particular, replace a local (non-gauge-invariant) $\tl{Z}$ operator with a non-local 
  Wilson operator consisting of product of the gauge fields. [Shown in Eq.~(\ref{eqn:gauss}) to (\ref{eqn:gaugedS}).]
  \item Examine how the composite swap operator~$S$ acts on spins and whether it 
  behaves as the mapping corresponding to gauging. The action of $S$ on some of spins may involve non-local charges, which is remedied by multiplying  $S$ with global charges. [Shown in Eq.~(\ref{cs1}) to (\ref{rep8}).]
\end{enumerate}
As we demonstrate below, there are exotic structures in the operators obtained by gauging spatially modulated symmetries, such as $0$-form and higher form
subsystem charges and algebraic relations between global and dipole charges.
\section{Subsystem symmetry}
\label{section3}
 In this section, we construct noninvertible operators obtained by gauging subsystem symmetry in two and three dimensions. 
\subsection{Two dimensions}
\begin{figure}
    \begin{center}
       \begin{subfigure}[h]{0.29\textwidth}
       \centering
  \includegraphics[width=0.8\textwidth]{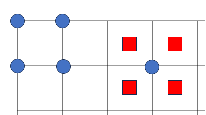}
         \caption{}\label{pl1}
             \end{subfigure}
             \hspace{5mm}
               \begin{subfigure}[h]{0.19\textwidth}
             \end{subfigure} 
                    \begin{subfigure}[h]
                    {0.39\textwidth}
                    \centering
  \includegraphics[width=0.8\textwidth]{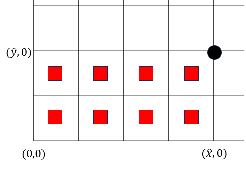}
         \caption{}\label{pl2}
             \end{subfigure}
             \hspace{5mm}
               \begin{subfigure}[h]{0.19\textwidth}
             \end{subfigure} 
 \end{center}
 \caption{(a)~(left) Plaquette Ising term defined in~\eqref{plaz}. (right)~The Gauss law, corresponding to~\eqref{gauss:pl}. The blue dots represent original spin degrees of freedom whereas the red squares do $\mathbf{Z}_2$~gauge fields. (b)~Example of~\eqref{membrane}. The black dot represents a node with coordinate $\br=(\hx,\hy).$}
 \end{figure}

Let us first concentrate on the two dimensions. To start, 
we introduce a 2D square lattice with the coordinate of each node being given by $\br\vcentcolon=(\hx,\hy)$, where $\hx$ and $\hy$ take integer numbers in the unit of lattice spacing. Also, we denote the coordinate of a plaquette as $\bp\vcentcolon=(\hx+\frac{1}{2},\hy+\frac{1}{2})$, and impose the periodic boundary condition on the lattice with system size $L_x\times L_y$. 
Defining two spin degrees of freedom on each node with Pauli operators $X_{\br}/Z_{\br}$, and $\tl{X}_{\br}/\tl{Z}_{\br}$,
we consider the following Hamiltonian composed of two copies of the plaquette Ising models:
\begin{eqnarray}
    H_{2D:plaquette}=-J\sum_{\bp}P_{Z,\bp}-h\sum_{\br}X_{\br}
    -\tl{J}\sum_{\bp}\tl{P}_{Z,\bp}-\tl{h}\sum_{\br}\tl{X}_{\br}.\label{pla}
\end{eqnarray}
Here, we have introduced 
operators consisting of four spins on corners on each plaquette, namely, 
\begin{eqnarray}
  {P}_{Z,\bp}\vcentcolon={Z}_{\bp+\ex+\ey}{Z}_{\bp+\ex-\ey}{Z}_{\bp-\ex+\ey}{Z}_{\bp-\ex-\ey}\nonumber\\
    \tl{P}_{Z,\bp}\vcentcolon=\tl{Z}_{\bp+\ex+\ey}\tl{Z}_{\bp+\ex-\ey}\tl{Z}_{\bp-\ex+\ey}\tl{Z}_{\bp-\ex-\ey}\label{plaz}
\end{eqnarray}
with vectors $\mathbf{e}_x\vcentcolon=(1,0)$, $\mathbf{e}_y\vcentcolon=(0,1)$ (See left term in Fig,~\ref{pl1}).
Note that due to the periodic boundary condition, 
$X_{\br+(L_x,0)}=X_{\br+(0,L_y)}=X_{\br}$ and the similar relation holds for other spin operators. 
\par
Similar to Sec.~\ref{section2}, we assume $J=\tl{J}$ and $h=\tl{h}$ in the following discussion. 
The Hamiltonian respects the following subsystem symmetries
\begin{eqnarray}
    Q_{sub_y,\hx}=\prod_{\hy=1}^{L_y}X_{\br},\;(1\leq \hx\leq L_x),\quad Q_{sub_x,\hy}=\prod_{\hx=1}^{L_x}X_{\br}\;(1\leq \hy\leq L_y)\nonumber\\
      \tl{Q}_{sub_y,\hx}=\prod_{\hy=1}^{L_y}\tl{X}_{\br},\;(1\leq \hx\leq L_x),\quad \tl{Q}_{sub_x,\hy}=\prod_{\hx=1}^{L_x}\tl{X}_{\br}\;(1\leq \hy\leq L_y),\label{19}
\end{eqnarray}
and the symmetry that exchanges between two spins at $\br$, namely, the Hamiltonian commutes with the swap operator:
\begin{eqnarray}
    S_{\br}=\frac{1}{2}(I+X_{\br}\tl{X}_{\br}+Z_{\br}\tl{Z}_{\br}-Z_{\br}\tl{Z}_{\br}X_{\br}\tl{X}_{\br})
\end{eqnarray}
The total swap operator is given by $S = \prod_{\hx=1}^{L_x}\prod_{\hy=1}^{L_y} S_{\br}$. Note that the symmetry algebra is the subsystem analog of the $D_8$ symmetry; along each horizontal or vertical line of the lattice, the model has the $D_8$ symmetry.  To wit, we have 
\begin{equation}
\begin{split}
    Q_{sub_y,\hx}^2= \tl{Q}_{sub_y,\hx}^2=S^2=1,\quad  SQ_{sub_y,\hx}= \tl{Q}_{sub_y,\hx}S,\quad
    S\tl{Q}_{sub_y,\hx}={Q}_{sub_y,\hx}S,\quad {Q}_{sub_y,\hx}\tl{Q}_{sub_y,\hx}=\tl{Q}_{sub_y,\hx}{Q}_{sub_y,\hx},
    \label{d8}
    \end{split}
\end{equation}
for $\forall \hx$ and similarly for other charges $Q_{sub_x,\hy}$,  $\tl{Q}_{sub_x,\hy}$ for $\forall \hy$.
The relation~\eqref{d8} is exactly the algebraic relation of $D_8$. 
The discussion presented in this subsection is valid as long as the model respects these symmetries~\eqref{d8} to the Hamiltonian~\eqref{pla}. One could add interacting terms to preserve these symmetries, examples of which have the form
\begin{eqnarray*}
    -g\sum_{\br}X_{\br}\tl{X}_{\br}-g^\prime\sum_{\bp}P_{Z,\bp}\tl{P}_{Z,\bp}.
\end{eqnarray*}
 \par
To proceed, we gauge the subsystem symmetries of the second Ising plaquette model~\cite{Vijay}. In doing so, we introduce extended Hilbert space on each plaquette of the 2D lattice whose Pauli operators are denoted as $\widehat{\tau}^X_{\br}$, $\widehat{\tau}^Z_{\br}$, corresponding to the gauge fields, 
and impose the following Gauss law:
\begin{eqnarray}
    \widehat{\tau}^X_{\br+\ex+\ey}   \widehat{\tau}^X_{\br+\ex-\ey}   \widehat{\tau}^X_{\br-\ex+\ey}   \widehat{\tau}^X_{\br-\ex-\ey}\tl{X}_{\br}=1,\label{gauss:pl}
\end{eqnarray}
which is depicted in right of Fig.~\ref{pl1}.
Intuition behind the form of the Gauss law~\eqref{gauss:pl} is that one decomposes global symmetries into local segments with introducing the extended Hilbert spaces (which are associated with the degrees of freedom of the gauge fields), corresponding to the fact that gauging is a procedure to promote global symmetries to local ones. Indeed, defining 
\begin{eqnarray*}
    \mathcal{G}_{sub,\br}\vcentcolon=    \widehat{\tau}^X_{\br+\ex+\ey}   \widehat{\tau}^X_{\br+\ex-\ey}   \widehat{\tau}^X_{\br-\ex+\ey}   \widehat{\tau}^X_{\br-\ex-\ey}\tl{X}_{\br}
\end{eqnarray*}
one has
\begin{eqnarray}
      \tl{Q}_{sub_y,\hx}=\prod_{\hy=1}^{L_y}\mathcal{G}_{sub,\br}\;(1\leq \hx\leq L_x),\quad \tl{Q}_{sub_x,\hy}=\prod_{\hx=1}^{L_x}\mathcal{G}_{sub,\br}\;(1\leq \hy\leq L_y).
\end{eqnarray}
The Gauss law is imposed by setting $\mathcal{G}_{sub,\br}=1\quad\forall\br$. In short, the Gauss law in Eq.~(\ref{gauss:pl}) is derived from the standard gauging procedure, i.e. promoting the global invariance into the local invariance and then
minimally coupling the system to background gauge fields which we turn to momentarily. 
\par
The four spins interaction defined on each plaquette is minimally coupled to the gauge field as
~\footnote{Regarding the original spin $\tl{Z}_{\br}$ as a matter field, this procedure is practically associated with subsystem analog of the minimally coupling to the gauge fields~\cite{Shavit_RevModPhys.52.453}.  } 
\begin{eqnarray}
   \tl{P}_{Z,\bp}\to \tl{P}_{Z,\bp}\widehat{\tau}^Z_{\bp}.
\end{eqnarray}
From the Gauss law~\eqref{gauss:pl}, 
and redefining $\tx_{\bp}\vcentcolon=\widehat{\tau}^X_{\bp}$, $\tz_{\bp}\vcentcolon=\tl{P}_{Z,\bp}\widehat{\tau}^Z_{\bp}$, 
we have the following mapping via gauging:
\begin{eqnarray}
    \tl{X}_{\br}\Rightarrow G_{X,\br},\quad  \tl{P}_{Z,\bp}\Rightarrow\tz_{\bp}
\end{eqnarray}
where $G_{X,\br}\vcentcolon=\tx_{\br+\ex+\ey}   \tx_{\br+\ex-\ey}   \tx_{\br-\ex+\ey}\tx_{\br-\ex-\ey}$.
The gauged Hamiltonian then becomes
\begin{eqnarray}
    \widehat{H}_{2D:plaquette}=-J\sum_{\bp}P_{\bp}-h\sum_{\br}X_{\br}
    -J\sum_{\br}\tz_{\bp}-h\sum_{\br}G_{X,\br},
\end{eqnarray}
Analogous to the previous argument presented around~\eqref{10} in the case of the double Ising chains, we expect the swap operator as the desired noninvertible operator. However, we have issue with a local term,~$\tl{Z}_{\br}$ inside the operator~$S_{\br}$
as it does not commute with the global symmetries~\eqref{19}. To fix this problem, we rewrite the term~$\tl{Z}_{\br}$ by a string of gauge fields. 
Setting the coordinate lattice so that $\br_0=(0,0)$ as the reference point, we assume $\tl{Z}_{\br}$ is transformed
into product of $\tau^Z$, forming a rectangular, whose four corners are given by nodes at $\br = (\hx,\hy)$ and $\br_0$, $(\hx,0)$, and $(0,\hy)$. More explicitly, we replace $\tl{Z}_{\br}$ with the following Wilson operator $W_{\br}$:
\begin{eqnarray}
   W_{\br}=\prod_{\hx^\prime=0}^{\hx-1}\prod_{\hy^\prime=0}^{\hy-1}(\tz_{\bp^\prime})\label{membrane}
\end{eqnarray}
with $\bp^\prime=(\hx^\prime+\frac{1}{2},\hy^\prime+\frac{1}{2})$.
We demonstrate one of the examples of~\eqref{membrane} in Fig,~\ref{pl2}.
Accordingly, the swap operator becomes
\begin{eqnarray}
    S_{\br}=\frac{1}{2}\left[(I+X_{\br}G_{\br})+Z_{\br}\prod_{\hx^\prime=0}^{\hx-1}\prod_{\hy^\prime=0}^{\hy-1}(\tz_{\bp^\prime})(I-X_{\br}G_{\br})\right].
\end{eqnarray}
One can verify that the swap operators $S_{\br}~(1\leq \hx\leq L_x-1,1\leq \hy\leq L_y-1)$ commute with themselves, yet they do not with $S_{(\hx,L_y)}$, $S_{(L_x,\hy)}$, $S_{(L_x,L_y)}$. Taking this fact into consideration, 
we define
\begin{eqnarray}
    S=\left(\prod_{\hx^\prime=1}^{L_x-1}\prod_{\hy^\prime=1}^{L_y-1}S_{\br^\prime}\right)\times \left(\prod_{\hx^\prime=1}^{L_x-1}S_{(\hx^\prime,L_y)}\times
    \prod_{\hy^\prime=1}^{L_y-1}S_{(L_x,\hy^\prime)}\times S_{(L_x,L_y)}
    \right).\label{swap1}
\end{eqnarray}
The product of the swap operators~\eqref{swap1} plays as the desired noninvertible operator for spins, except the ones in vicinity of the end points of the lattice. Indeed, after some algebra, one finds that 
\begin{eqnarray}
SG_{X,\br}=X_{\br}S,\quad
    SX_{\br}=\begin{cases}G_{X,\br}S\quad (1\leq \hx\leq L_x-1, 1\leq \hy\leq L_y-1)\\
   G_{X,\br}SQ_{sub_x,\hy} \quad (\hx=L_x, 1\leq \hy\leq L_y-1)\\
         G_{X,\br}SQ_{sub_y,\hx} \quad (1\leq\hx\leq L_x-1, \hy=L_y)\\
         G_{X,\br}SQ_{sub_x,\hy=L_y}Q_{sub_y,\hx=L_x}Q_{all}\quad (\hx=L_x,\hy=L_y)
    \end{cases}\label{16}
\end{eqnarray}
where $Q_{all}=\prod_{\hx=1}^{L_x}Q_{sub_y,\hx}=\prod_{\hy=1}^{L_y}Q_{sub_x,\hy}$. 
Indeed, the first line can be checked directly as
\begin{equation}
\begin{split}
S_{\br} X_{\br} &= \frac{1}{2} \left[ (X_{\br} + G_{\br}) + Z_{\br} \prod_{\hx^\prime=0}^{\hx-1}\prod_{\hy^\prime=0}^{\hy-1}(\tz_{\bp^\prime}) (X_{\br} - G_{\br}) \right] \\
&= \frac{1}{2} \left[ (X_{\br} + G_{\br}) - Z_{\br} \prod_{\hx^\prime=0}^{\hx-1}\prod_{\hy^\prime=0}^{\hy-1}(\tz_{\bp^\prime}) G_{\br} (1-X_{\br} G_{\br}) \right] \\
&=  \frac{1}{2} \left[ (X_{\br} + G_{\br}) + G_{\br} Z_{\br} \prod_{\hx^\prime=0}^{\hx-1}\prod_{\hy^\prime=0}^{\hy-1}(\tz_{\bp^\prime})  (1-X_{\br} G_{\br}) \right] \\
&= G_{\br}  \frac{1}{2} \left[ (G_{\br} X_{\br} +1) + Z_{\br} \prod_{\hx^\prime=0}^{\hx-1}\prod_{\hy^\prime=0}^{\hy-1}(\tz_{\bp^\prime}) G_{\br} (1-X_{\br} G_{\br}) \right] \\
&= G_{\br} S_{\br}.
\end{split}
\end{equation}
and other relations can be derived analogously. 
Also, we have 
\begin{eqnarray}
    SP_{Z,\bp}=\begin{cases}\tz_{\bp}S\quad (1\leq \hx\leq L_x-1, 1\leq \hy\leq L_y-1)\\
   \tz_{\bp}S\eta_{sub_x,\hy} \quad (\hx=0, 1\leq \hy\leq L_y-1)\\
         \tz_{\bp}S\eta_{sub_y,\hx} \quad (1\leq\hx\leq L_x-1, \hy=0)\\
         \tz_{\bp}S\eta_{sub_x,\hy}\eta_{sub_y,\hx}\eta_{all}\quad (\hx=0,\hy=0),
    \end{cases}
S\tz_{\bp}=\begin{cases}P_{Z,\bp}S\quad (1\leq \hx\leq L_x-1, 1\leq \hy\leq L_y-1)\\
  P_{Z,\bp}S\eta_{sub_x,\hy} \quad (\hx=0, 1\leq \hy\leq L_y-1)\\
       P_{Z,\bp}S\eta_{sub_y,\hx} \quad (1\leq\hx\leq L_x-1, \hy=0)\\
         P_{Z,\bp}S\eta_{sub_x,\hy}\eta_{sub_y,\hx}\eta_{all}\quad (\hx=0,\hy=0).
    \end{cases}\label{17}
\end{eqnarray}
Here, 
\begin{eqnarray}
    \eta_{sub_x,\hy}\vcentcolon=\prod_{\hx=0}^{L_x-1}\tz_{\bp},\quad  \eta_{sub_y,\hx}\vcentcolon=\prod_{\hy=0}^{L_y-1}\tz_{\bp}
\end{eqnarray}
which are charges associated with subsystem symmetries that have emerged after gauging~\footnote{They are subsystem analog of the \textit{quantum symmetries}~\cite{Vafa1989quantumsym}.} and 
\begin{equation*}
    \eta_{all}=\prod_{\hx=0}^{L_x-1}\eta_{sub_y,\hx}=\prod_{\hy=0}^{L_y-1}\eta_{sub_x,\hy}.
\end{equation*}
\par
Relations~\eqref{16} and~\eqref{17} indicate that 
around the end points, 
the operator $S$ acts as a mapping corresponding to gauging, \textit{up to} 
charges of subsystem symmetries, such as $Q_{sub_x,\hy}$. 
To fix this issue, 
a proper operator is defined by multiplying projection operators with $S$, namely, we introduce the following:
\begin{eqnarray}
    D\vcentcolon=\frac{1}{2^{L_x+L_y-1}}S\times\left[\frac{1}{2}\prod_{\hy=1}^{L_y}(1+Q_{sub_x,\hy})\times\prod_{\hx=1}^{L_x}(1+Q_{sub_y,\hx})\right]\times\left[\frac{1}{2}\prod_{\hy=0}^{L_y-1}(1+\eta_{sub_x,\hy})\times\prod_{\hx=0}^{L_x-1}(1+\eta_{sub_y,\hx})\right].
\end{eqnarray}
From~\eqref{16}~and~\eqref{17}, it is immediate to check that 
\begin{eqnarray}
 DG_{X,\br}=X_{\br}D,\quad   DX_{\br}=G_{X,\br}D,\quad DP_{Z,\bp}=\tz_{\bp}D,\quad D\tz_{\bp}=P_{Z,\bp}D,
\end{eqnarray}
implying that the operator $D$ is the desired swap operator. 
Further, this operator is noninvertible; the fusion rules of the operators read
\begin{eqnarray}
    D\times D&=&\left[\frac{1}{2}\prod_{\hy=1}^{L_y}(1+Q_{sub_x,\hy})\times\prod_{\hx=1}^{L_x}(1+Q_{sub_y,\hx})\right]\times\left[\frac{1}{2}\prod_{\hy=0}^{L_y-1}(1+\eta_{sub_x,\hy})\times\prod_{\hx=0}^{L_x-1}(1+\eta_{sub_y,\hx})\right],\nonumber\\
    \xi D&=&D\xi=D\quad (\xi=Q_{sub_x,\hy},Q_{sub_y,\hx},\eta_{sub_x,\hy},\eta_{sub_y,\hx}~\forall~\hx,\hy).\label{33}
\end{eqnarray}
Hence, we have constructed subsystem analog of the $\text{Rep}(D_8)$ noninvertible operators~\eqref{rep8}. Here, the fusion rule of the operators yield subsystem charges.

\begin{figure}
    \begin{center}
       \begin{subfigure}[h]{0.49\textwidth}
       \centering
  \includegraphics[width=0.8\textwidth]{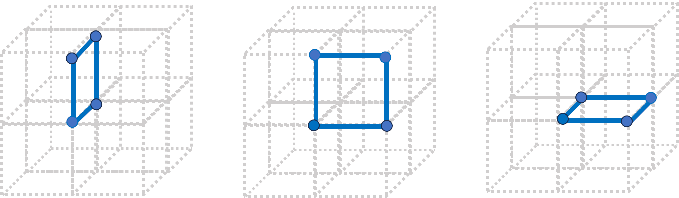}
         \caption{}\label{isingpl1}
             \end{subfigure}
             \hspace{5mm}
                   \begin{subfigure}[h]{0.20\textwidth}
                   \centering
 \includegraphics[width=1.5\textwidth]{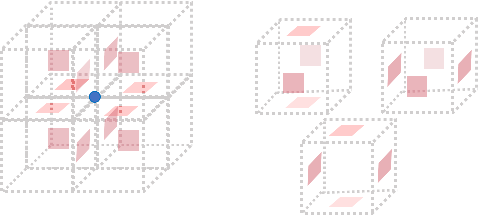}
         \caption{}\label{gauss}
             \end{subfigure}
             \hspace{5mm}
               \begin{subfigure}[h]{0.19\textwidth}
             \end{subfigure} 
       \begin{subfigure}[h]{0.6\textwidth}
       \centering
  \includegraphics[width=1\textwidth]{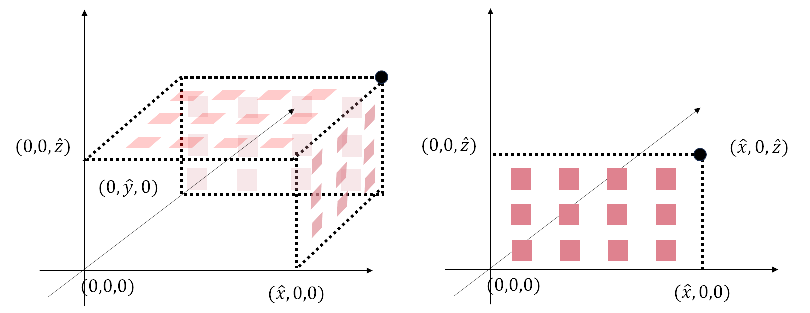}
         \caption{}\label{disk0}
             \end{subfigure}
      \end{center}
 \caption{(a)~Three types of plaquette Ising terms that constitute the Hamiltonian~\eqref{3D_Ising}. (b) The Gauss law~\eqref{gs} is described by the left configuration whereas the three flux operators~\eqref{flux} are depicted by the configurations on the right.
 The original spin degrees of freedom are depicted as blue dots whereas the gauge fields are indicated by red rectangles.
 (c) Configuration of an operator given in~\eqref{Mr}.}
 \end{figure}
\subsection{Three dimensions}
In this subsection, we turn to construction of the operators via gauging subsystem symmetries in the
3D Ising plaquette model.\par
To facilitate the following discussion, we introduce several notations. We define coordinate of the node of the 3D cubic lattice as~$\br=(\hx,\hy,\hz)$. Also, we denote coordinate of a plaquette on $xy$, $yz$, $zx$-plane as $\bp_{xy}=(\hx+\frac{1}{2},\hy+\frac{1}{2},\hz)$, $\bp_{yz}=(\hx,\hy+\frac{1}{2},\hz+\frac{1}{2})$, $\bp_{zx}=(\hx+\frac{1}{2},\hy,\hz+\frac{1}{2})$, respectively. For latter purposes, we introduce vectors $\exx=(1,0,0)$, $\eyy=(0,1,0)$, $\ezz=(0,0,1)$ and coordinate of a cube $\bc=(\hx+\frac{1}{2},\hy+\frac{1}{2},\hz+\frac{1}{2})$.
We think of a 3D cubic lattice and define two spin degrees of freedom on each node with Pauli operators $X_{\br}/Z_{\br}$,~and~$\tl{X}_{\br}/\tl{Z}_{\br}$. 
We impose the periodic boundary condition with the system size $L_x\times L_y\times L_z$.\par With these preparations, 
we introduce the following Hamiltonian: 
\begin{eqnarray}
    H_{3D:plaquette}=-J\sum_{\bp_{ab}}P_{Z,\bp_{ab}}-h\sum_{\br}X_{\br}-\tl{J}\sum_{\bp_{ab}}\tl{P}_{Z,\bp_{ab}}-\tl{h}\sum_{\br}\tl{X}_{\br},\label{3D_Ising}
\end{eqnarray}
where $\sum_{\bp_{ab}}$ denotes summing over all plaquettes in the cubic lattice (more precisely $\bp_{ab}=\bp_{xy},\bp_{yz},\bp_{zx}$ ) and  
\begin{align}
    P_{Z,\bp_{ab}}&\vcentcolon=Z_{\bp_{ab}-\ea-\eb}Z_{\bp_{ab}-\ea+\eb}Z_{\bp_{ab}+\ea+\eb}Z_{\bp_{ab}+\ea-\eb}
\end{align}
See also Fig.~\ref{isingpl1}.
The term $\tl{P}_{Z,\bp_{ab}}$ is similarly defined by replacing $Z_{\br}$ with $\tl{Z}_{\br}$. 
 In what follows, we set $J=\tl{J}$ and $h=\tl{h}$. 
 In such a case, the model~\eqref{3D_Ising} respects the following subsystem symmetries~(i.e., global spin flip on a plane):
 \begin{align}
     Q_{sub_{ab},\hat{c}} &= \prod_{\hat{a}=1}^{L_a}\prod_{\hat{b}=1}^{L_b}X_{\br}\quad(1\leq \hat{c} \leq L_c),\label{sub3}
 \end{align}
 where $a,b,c$ are cyclic permutations of $x,y,z$. Likewise,  
$\tl{Q}_{sub_{xy},\hz}$, $\tl{Q}_{sub_{yz},\hx}$, $\tl{Q}_{sub_{zx},\hy}$ are defined by replacing~$X_{\br}$ with~$\tl{X}_{\br}$ in~\eqref{sub3}, and the $\mathbb{Z}_2$ symmetry, exchanging local spins without tilde and the ones with tilde. To wit, the model~\eqref{3D_Ising} commutes with an operator $S$, where
\begin{eqnarray}
    S=\prod_{\hx=1}^{L_x}\prod_{\hy=1}^{L_y}\prod_{\hz=1}^{L_z}S_{\br},\quad  S_{\br}=\frac{1}{2}(I+X_{\br}\tl{X}_{\br}+Z_{\br}\tl{Z}_{\br}-Z_{\br}\tl{Z}_{\br}X_{\br}\tl{X}_{\br}).\label{29}
\end{eqnarray}
Similar to~\eqref{d8} in the previous subsection, the model~\eqref{3D_Ising} admits the analogous $D_8$ subsystem symmetry.
Notice that discussion in what follows is valid as long as the model we consider respects the $D_8$ subsystem symmetry~\eqref{29}. 
\par
Now we construct noninvertible operator operator by gauging subsystem symmetries for spins~$\tl{Z}_{\br}$.~\footnote{The procedure of gauging such symmetries in 3D was discussed in~\cite{Vijay}.} To this end, we accommodate the extended Hilbert space on each plaquette with Pauli operators represented by $\tl{\tau}^Z$ and $\tl{\tau}^X$. The Gauss law reads (see also left term in Fig.~\ref{gauss})
\begin{eqnarray}
    \tl{X}_{\br}\times\prod_{\partial \bp_{xy}\in \br}\tl{\tau}^X_{\bp_{xy}}\times \prod_{\partial \bp_{yz}\in \br}\tl{\tau}^X_{\bp_{yz}}\times \prod_{\partial \bp_{zx}\in \br}\tl{\tau}^X_{\bp_{zx}}=1.\label{gs}
\end{eqnarray}
Here, $\partial \bp_{ab}\in \br$ denotes plaquettes on $ab$-plane sharing a corner with a node at $\br$. 
Similar to the previous subsection, intuitive understanding of the Gauss law term~\eqref{gs} is that one crops the global charges into local ones; defining 
\begin{eqnarray*}
    \mathcal{G}_{sub3D,\br}\vcentcolon=  \tl{X}_{\br}\times\prod_{\partial \bp_{xy}\in \br}\tl{\tau}^X_{\bp_{xy}}\times \prod_{\partial \bp_{yz}\in \br}\tl{\tau}^X_{\bp_{yz}}\times \prod_{\partial \bp_{zx}\in \br}\tl{\tau}^X_{\bp_{zx}},
\end{eqnarray*}
it follows that 
\begin{eqnarray}
     Q_{sub_{ab},\hat{c}}=\prod_{\hat{a}=1}^{L_a}\prod_{\hat{b}=1}^{L_b}\mathcal{G}_{sub3D,\br}\quad(1\leq\hat{c}\leq L_c),
\end{eqnarray}
where $a,b,c$ are cyclic permutations of $x,y,z$.
The Gauss law is imposed by setting $\mathcal{G}_{sub3D,\br}=1\quad\forall\br$.\par
Plaquette Ising terms~$\tl{P}_{\bp_{ab}}$ 
are minimally coupled to the gauge field as
\begin{eqnarray}
    \tl{P}_{\bp_{ab}}\to  \tl{P}_{\bp_{ab}}\tl{\tau}^Z_{\bp_{ab}}.
\end{eqnarray}
Defining new variables as $\tx_{\bp_{ab}}\vcentcolon=\tl{\tau}^X_{\bp_{ab}}$, $\tz_{\bp_{ab}}\vcentcolon=\tl{P}_{\bp_{ab}}\tl{\tau}^Z_{\bp_{ab}}$, we obtain the following mapping via gauging:
\begin{eqnarray}
    \tl{X}_{\br}\Rightarrow  G_{X,\br},\quad \tl{P}_{\bp_{ab}}\Rightarrow \tau^Z_{\bp_{ab}}\label{map}
\end{eqnarray}
where
\begin{eqnarray}
    G_{X,\br}\vcentcolon=\prod_{\partial \bp_{xy}\in \br}{\tau}^X_{\bp_{xy}}\times \prod_{\partial \bp_{yz}\in \br}{\tau}^X_{\bp_{yz}}\times \prod_{\partial \bp_{zx}\in \br}{\tau}^X_{\bp_{zx}}.
\end{eqnarray}
We also
add the following operators to the Hamiltonian
\begin{equation}
\begin{split}
  &-g_x\sum_{\bc}B_{\bc,x}-g_y\sum_{\bc}B_{\bc,y}-g_z\sum_{\bc}B_{\bc,z} \vcentcolon= \\
  &-g_{x}\sum_{\bc}\prod_{\bp_{xy}\in\partial\bc}\tz_{\bp_{xy}}\times \prod_{\bp_{zx}\in\partial\bc}\tz_{\bp_{zx}}-g_{y}\sum_{\bc}\prod_{\bp_{yz}\in\partial\bc}\tz_{\bp_{yz}}\times \prod_{\bp_{xy}\in\partial\bc}\tz_{\bp_{xy}}-g_{z}\sum_{\bc}\prod_{\bp_{zx}\in\partial\bc}\tz_{\bp_{zx}}\times \prod_{\bp_{yz}\in\partial\bc}\tz_{\bp_{yz}},
   \quad(g_x,g_y,g_z>0)\label{flux}
   \end{split}
\end{equation}
to ensure that the theory becomes dynamically trivial~(See also right configurations in Fig.~\ref{gauss}.) which amounts to imposing flatness condition of the gauge fields on the theory. 
Here,~$\prod_{\bp_{ab}\in\partial\bc}$ denotes product of the gauge fields over plaquettes on $ab$-plane that surround the cube~$\bc$.
After gauging, we finally arrive at the following gauged Hamiltonian:
\begin{eqnarray}
    \widehat{H}_{3D:plaquette}=&-&J\sum_{\bp_{ab}}P_{Z,\bp_{ab}}-h\sum_{\br}X_{\br}-{J}\sum_{\bp_{ab}}{\tz}_{\bp_{ab}}\nonumber\\
    &-&{h}\sum_{\br}G_{X,\br}-g_x\sum_{\bc}B_{\bc,x}-g_y\sum_{\bc}B_{\bc,y}-g_z\sum_{\bc}B_{\bc,z}. \label{3dpl}
\end{eqnarray}
Note that the second line in~\eqref{3dpl} is nothing but the Hamiltonian of the X-cube model~\cite{Vijay}.~\footnote{Note that 
three flux operators are not independent. Indeed, 
one of them, say, $B_{\bc,z}$ is generated by other flux terms,  $B_{\bc,x}$ and $B_{\bc,y}$ via $B_{\bc,z}=B_{\bc,x}B_{\bc,y}$.}
\par
The operator is constructed from the swap operator after gauging. Similar to the argument in the previous subsection, 
a local spin operator $\tl{Z}_{\br}$
in the swap operator~\eqref{29} does not commute with the subsystem symmetries, $\tl{Q}_{ab}$, hence, it should become non-local operators after the gauging.
We proceed by replacing  $\tl{Z}_{\br}$ with the Wilson operator $\mathcal{M}_{\br}$ comprised of product of the gauge fields, forming membranes, that is, 
\begin{eqnarray}
 \mathcal{M}_{\br}=
    \begin{cases}
\left(\prod_{\hy^\prime=0}^{\hy}\prod_{\hz^\prime=0}^{\hz}\tz_{\bp_{yz}^\prime}\right)\times \left(\prod_{\hz^\prime=0}^{\hz}\prod_{\hx^\prime=0}^{\hx}\tz_{\bp_{zx}^\prime}\right)\times \left(\prod_{\hx^\prime=0}^{\hx}\prod_{\hy^\prime=0}^{\hy}\tz_{\bp_{xy}^\prime}\right)~(x,y,z\neq 0)\\
        \prod_{\hat{a}^\prime=0}^{\hat{a}} \prod_{\hat{b}^\prime=0}^{\hat{b}}\tau^Z_{\bp_{ab}}~(\text{When}~\bp_{ab}\text{is on a}~$ab-$\text{plane that intersects the origin.})
    \end{cases}.\label{Mr}
\end{eqnarray}
Such configurations are portrayed in Fig.~\ref{disk0}.
\begin{figure}[t]
    \begin{center}
        
       \includegraphics[width=1\textwidth]{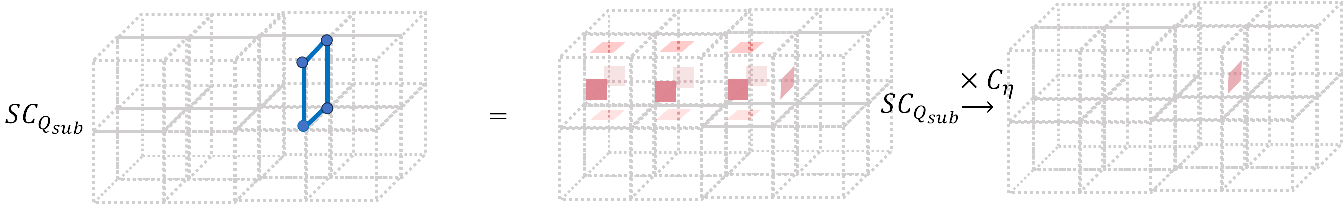}
\end{center}
       \caption{Pictorial understanding of how non-local terms in~\eqref{br} is absorbed in the operator $C_{\eta}$~\eqref{eeta}.
 }\label{fig1}

   \end{figure}
The swap operator on each node is described by
\begin{eqnarray}
    S_{\br}=
    \begin{cases}
      \frac{1}{2}\left[I+X_{\br}G_{X,\br}+Z_{\br}(I-X_{\br}G_{X,\br})\right]~[\br=(0,0,0),(\hx,0,0),(0,\hy,0),(0,0,\hz),\hx,\hy,\hz\neq 0]\\
          \frac{1}{2}\left[I+X_{\br}G_{X,\br}+Z_{\br}\mathcal{M}_{\br}(I-X_{\br}G_{X,\br})\right]~(\text{else})
    \end{cases}
  .\label{sr}
\end{eqnarray}
Note that $S_{\br}$ with the first (second) relation in~\eqref{sr}
is invertible (noninvertible) by noticing that $S_{\br}^2=1~(S_{\br}^2\neq 1)$. The operator $S_{(0,0,0)}$ does not commute with $S_{\br}$ with $\br$ being on $xy$-, $yz$-, $zx$-planes that intersect the origin. Also, $S_{(\hx,0,0)}$ does not commute with $S_{\br}$ on $yz$-plane that intersects $x$-axis with the coordinate $(\hx,0,0)$, and the similar relation holds when permuting $x$, $y$, and $z$.
The operator $S$ is noninvertible due to the fact that $S_{\br}^2\neq 1$ in the second case of~\eqref{sr}.
\par
We define the product of the swap operator as
\begin{eqnarray}
    S=\left(\prod_{\hx=1}^{L_x-1}S_{(\hx,0,0)}\times\prod_{\hy=1}^{L_y-1}S_{(0,\hy,0)}\times\prod_{\hz=1}^{L_z-1}S_{(0,0,\hz)}\times S_{(0,0,0)}\right)\times \left(\prod_{\hx=1}^{L_x-1}\prod_{\hy=1}^{L_y-1}\prod_{\hz=1}^{L_z-1}S_{\br}\right)
\end{eqnarray}
Here, the order of operator is important as the terms in the first braket does not commute with some of the terms in the second braket. 

Now we investigate how the operator $S$ acts on a local spin $X_{\br}$ and $G^X_{\br}$.
 After some algebra, one finds 
\begin{eqnarray}
SX_{(0,0,0)}&=&G^X_{(0,0,0)}SQ_{sub_{xy},\hz=0}Q_{sub_{yz},\hx=0}Q_{sub_{zx},\hy=0},\nonumber\\
SX_{(\hx,0,0)}=G^X_{(\hx,0,0)}SQ_{sub_{yz},\hx},\quad SX_{(0,\hy,0)}&=&G^X_{(0,\hy,0)}SQ_{sub_{zx},\hy},\quad SX_{(0,0,\hz)}=G^X_{(0,0,\hz)}SQ_{sub_{xy},\hz},\nonumber\\
SX_{\br}&=&G^X_{\br}S~[\br\neq(0,0,0),(\hx,0,0),(0,\hy,0),(0,0,\hz)],\nonumber\\
SG^X_{\br}&=&X_{\br}S~\forall\br.
\end{eqnarray}
As seen from the last two relations, the points at the origin and three spatial axes, the operator $S$ plays the role of the operators implementing the first mapping in~\eqref{map} whereas acting $S$ on the spin around the boundary involves subsystem global charges.
To fix the issue, one multiplies subsystem charges with~$S$, namely, $S$ is modified as $SC_{Q_{sub}}$,
where
\begin{eqnarray}
    C_{Q_{sub}}\vcentcolon=\frac{1}{2^2}\prod_{\hz=1}^{L_z}\left(1+Q_{sub_{xy},\hz}\right)\times \prod_{\hx=1}^{L_x}\left(1+Q_{sub_{yz},\hx}\right)\times \prod_{\hy=1}^{L_y}\left(1+Q_{sub_{zx},\hy}\right).
\end{eqnarray}
Here, the prefactor $\frac{1}{2^2}$ is introduced to take care of redundancy due to the relation
\begin{equation*}
\prod_{\hz=1}^{L_z}Q_{sub_{xy},\hz}=\prod_{\hx=1}^{L_x}Q_{sub_{yz},\hx}=\prod_{\hy=1}^{L_y}Q_{sub_{zx},\hy}.
\end{equation*}
With this modification, one has
\begin{equation*}
    SC_{Q_{sub}}X_{\br}=G^X_{\br}SC_{Q_{sub}}~[\br=(0,0,0),(\hx,0,0),(0,\hy,0),(0,0,\hz)],
\end{equation*}
implying $SC_{Q_{sub}}$ is the operator which maps $X_{\br}$ to $G^X_{\br}$ and vice versa, reproducing the first mapping in~\eqref{map}. \par
We also investigate how the operator $SC_{Q_{sub}}$ acts on other spin coupling terms and $\tau^Z_{\bp_{ab}}$. For instance, focusing on a coupling term $P_{\bp_{yz}}~(\hx\neq 0)$, one has
\begin{eqnarray}
    SC_{Q_{sub}}P_{\bp_{yz}}=\tau^Z_{\bp_{yz}}\times\left[\left(\prod_{\hx^\prime=0}^{\hx}\prod_{\hz^\prime=\hz}^{\hz+1}\tau^Z_{\bp^\prime_{xy}}\right)\times\left(\prod_{\hx^\prime=0}^{\hx}\prod_{\hy^\prime=\hy}^{\hy+1}\tau^Z_{\bp^\prime_{yz}}\right)\right]SC_{Q_{sub}},\label{br}
\end{eqnarray}
which is portrayed in the middle of Fig.~\ref{fig1}. This indicates that when $SC_{Q_{sub}}$ acts on $P_{\bp_{yz}}~(\hx\neq 0)$, it picks up nonlocal operators, corresponding to the argument inside the braket $[*]$ in~\eqref{br}. To remedy this issue, we introduce the following operator:
\begin{eqnarray}
    C_{\eta}\vcentcolon=\frac{1}{2^{2L_xL_yL_z}}\sum_{\gamma}\eta(\gamma),\label{eeta}
\end{eqnarray}
where $\sum_{\gamma}\eta(\gamma)$ stands for summing over all kinds of closed loops of the gauge fields, $\tau^Z_{\bp_{ab}}$, including noncontractible ones.

With this $C_{\eta}$, we defined the following operator:
\begin{eqnarray}
     D\vcentcolon=\frac{1}{2^{L_x+L_y+L_z-2}}S\times C,\label{sc}
 \end{eqnarray}
 where $C\vcentcolon=C_{Q_{sub}}C_{\eta}$. From~\eqref{flux}, and noticing that the argument inside the square bracket of Eq.~\eqref{br} can be written as the product of $B_{\bc,x}$'s, it follows that such an argument is absorbed into $C_{\eta}$ [See right configuration of Fig.~\ref{fig1}.]. Therefore, we have
 \begin{eqnarray}
     DP_{\bp_{yz}}=\tau^Z_{\bp_{yz}}D
 \end{eqnarray}
reproducing the desired second mapping in~\eqref{map}. 
The similar statement holds true for other terms, $P_{\bp_{yz}}$, and $P_{\bp_{zx}}$. Further,  
the analogous line of thoughts leads to that 
\begin{eqnarray}
     D\tau^Z_{\bp_{ab}}=P_{\bp_{ab}}D.
\end{eqnarray}
Note that when taking the limit $g_x,g_y,g_z\to\infty$, $C_{\eta}$ becomes (see Appendix.~\ref{app:1} for more explanations)
\begin{eqnarray}
    C_{\eta}&=&\frac{1}{2^{L_x+L_y+L_z-1}}\left[\frac{1}{2}\prod_{\hz=1}^{L_z}\left(1+\eta^{zx}_x(\hz)\right)\left(1+\eta^{yz}_y(\hz)\right)\right]\times \left[\frac{1}{2}\prod_{\hx=1}^{L_x}\left(1+\eta^{zx}_z(\hx)\right)\left(1+\eta^{xy}_y(\hx)\right)\right]\nonumber\\
 &&\times \left[\frac{1}{2}\prod_{\hy=1}^{L_y}\left(1+\eta^{xy}_x(\hy)\right)\left(1+\eta^{yz}_z(\hy)\right)\right],\label{cond}
\end{eqnarray}
where
 \begin{eqnarray}
 \eta_{a}^{ab}(\hat{b})\vcentcolon=\prod_{\hat{a}=1}^{L_a}\tz_{\bp_{ab}},\quad  \eta_{b}^{ab}(\hat{a})\vcentcolon=\prod_{\hat{b}=1}^{L_b}\tz_{\bp_{ab}},\label{ab}
 \end{eqnarray}
 which corresponds to noncontractible of the gauge fields, associated with lineon excitations of the $X$-cube model.~\footnote{The prefactor $\frac{1}{2}$ inside the bracket in~\eqref{cond} is introduced so that it takes care of the redundancy due to the relation $\prod_{\hat{b}=1}^{L_b}\eta^{ab}_a(\hat{b})=\prod_{\hat{a}=1}^{L_a}\eta^{ab}_b(\hat{a})$. } 
 As an example, $\eta^{xy}_x(\hat{y})$ [$\eta^{xy}_y(\hat{x})$] denotes a non-contractible loop of the lineon in the $x$[$y$]-direction formed by gauge fields on a $xy$-plane at the coordinate $\hat{y}[\hx]$. Note that these loops do not have $\hz$ dependence since they are topological in the $z$-direction. 
 The operator in~\eqref{cond} is the subsystem analog of the condensation defects in continuum limit~\cite{Roumpedakis:2022aik}.~\footnote{See also~\cite{Bucher:1991bc,Chesire_Nayak} for earlier expositions on the related topic.}\par
 To summarize the argument, we have constructed operator $D$~\eqref{sc}, 
 satisfying
 \begin{eqnarray}
     DX_{\br}=G^X_{\br}D,\quad  DG^X_{\br}=X_{\br}D,\quad DP_{\bp_{ab}}=\tau^Z_{\bp_{ab}}D,\quad \tau^Z_{\bp_{ab}}D=DP_{\bp_{ab}}.
 \end{eqnarray}
 This operator $D$ is noninvertible. Indeed, we obtain the following fusion rule:
 \begin{eqnarray}
     D\times D=C,\quad C\times C=2^{2(L_x+L_y+L_z)-4}C,\quad DC=CD=2^{2(L_x+L_y+L_z)-4}D.\label{57}
 \end{eqnarray}
 In particular, in the limit $g_x,g_y,g_z\to\infty$, $C$ becomes
 \begin{equation}
 \begin{split}
     C=&\left\{\frac{1}{2^2}\prod_{\hz=1}^{L_z}\left(1+Q_{sub_{xy},\hz}\right)\times \prod_{\hx=1}^{L_x}\left(1+Q_{sub_{yz},\hx}\right)\times \prod_{\hy=1}^{L_y}\left(1+Q_{sub_{zx},\hy}\right)\right\} \\
    & \times \frac{1}{2^{L_x+L_y+L_z-1}}\left[\frac{1}{2}\prod_{\hz=1}^{L_z}\left(1+\eta^{zx}_x(\hz)\right)\left(1+\eta^{yz}_y(\hz)\right)\right]\times \left[\frac{1}{2}\prod_{\hx=1}^{L_x}\left(1+\eta^{zx}_z(\hx)\right)\left(1+\eta^{xy}_y(\hx)\right)\right]\\
    & \times \left[\frac{1}{2}\prod_{\hy=1}^{L_y}\left(1+\eta^{xy}_x(\hy)\right)\left(1+\eta^{yz}_z(\hy)\right)\right],
    \end{split}\label{58}
 \end{equation}
 The operator $D$ constructed in this subsection is the subsystem analog of 
$2$-Rep$\left((\mathbb{Z}_2^{(1)}\times\mathbb{Z}_2^{(1)})\rtimes\mathbb{Z}_2^{(0)}\right)$ studied in the fusion $2$-category theory~\cite{bhardwaj2023non,Choi:2024rjm}. Here, $2$-Rep$\left((\mathbb{Z}_2^{(1)}\times\mathbb{Z}_2^{(1)})\rtimes\mathbb{Z}_2^{(0)}\right)$ 
stands for the $2$-representation of the $2$-group~$(\mathbb{Z}_2^{(1)}\times\mathbb{Z}_2^{(1)})\rtimes\mathbb{Z}_2^{(0)}$, where $\mathbb{Z}_2^{(1)}$ denotes $1$-form symmetry and $\mathbb{Z}_2^{(0)}$ does 
$0$-form symmetry exchanging two $1$-form symmetries. 
In our case, 
the fusion rule involves $0$-form subsystem charges and higher form operators corresponding to the lineon excitations found in the $X$-cube model. \par

We conclude this section with a comment that the noninvertible operators $D$ for subsystem symmetries discussed in this section and dipole symmetries discussed in Sec.~\ref{section4} are translational invariant and Hermitian, which follows from similar arguments in Appendix~E of Ref.~\cite{Choi:2024rjm}.

\section{Dipole symmetry}
\label{section4}
In this section, we turn to the case of the dipole symmetry in one and two dimensions. As we will demonstrate, the fusion rules of the operators are characterized by dipole algebra, which is a new structure that intertwines global and dipole charges.
In what follows, we consistently use ``$0$-form dipole symmetry'' to refer to the two $\mathbb{Z}_N$ generators as global $\mathbb{Z}_N$ uniform and dipole symmetries. 
\subsection{One dimension}
\label{subsec:1ddipole}
In this subsection, we study the operator
in the $\mathbb{Z}_N$ chain model respecting $0$-form dipole symmetry. To do so, we introduce several notations. We define $Z_j$ and $X_j$ by the clock and shift matrices at site index~$j$, respectively. The commutation relation for onsite spins satisfies $Z_j X_{j} = \omega X_{j} Z_j$, where $\omega = e^{2\pi i /N}$.
Also, denoting the system size as $L$, we impose a periodic boundary condition so that $X_{j=0}=X_{j=L}$ and for simplicity, we
assume $L=kN\;(k\in\mathbb{Z})$. \par
To set the stage, 
we think of the following two copies of the spin chains with the $0$-form dipole symmetries:
\begin{eqnarray}
    H_{1D:dipole}=-J\sum_j Z_{j-1}(Z_j^\dagger)^2Z_{j+1}-h\sum_jX_j-\tl{J}\sum_j \tl{Z}_{j-1}(\tl{Z}_j^\dagger)^2\tl{Z}_{j+1}-\tl{h}\sum_j\tl{X}_j+h.c.\label{eq:spin}
\end{eqnarray}
In the following, we set $J=\tl{J}$ and $h=\tl{h}$.
The Hamiltonian~\eqref{eq:spin} respects the following symmetries:
\begin{eqnarray}
    Q_{0}=\prod_{j=1}^LX_j,\quad  Q_{dipole}=\prod_{j=1}^L(X_j)^j\nonumber\\
      \tl{Q}_{0}=\prod_{j=1}^L \tl{X}_j,\quad  \tl{Q}_{dipole}=\prod_{j=1}^L(\tl{X}_j)^j.
    \label{dip}
\end{eqnarray}
 The charges~\eqref{dip} are the typical examples of $0$-form dipole symmetry. The charges~\eqref{dip}
 satisfy~\footnote{Notice that taking $N=2$ reduces the $\mathbb{Z}_2 \times \mathbb{Z}^{dipole}_2$ symmetry into regular $\mathbb{Z}_2 \times \mathbb{Z}_2$ symmetry for even and odd sublattices.}
\begin{eqnarray}
 TQ_{dipole}T^{-1}=Q_0^\dagger Q_{dipole}\label{64}
\end{eqnarray}
and similarly for $\tl{Q}_{dipole}$. 
Here, $T$ represents translational operator, shifting one lattice constant. More explicitly, $TX_jT^{-1}=X_{j+1}$.
The model~\eqref{eq:spin} also respects $\mathbb{Z}_2$ symmetry exchanging spins. To see this, 
we introduce
the generalized swap operator,
\begin{eqnarray}
    S_j\vcentcolon=\frac{1}{N}\sum_{\alpha,\beta=1}^N\omega^{-\alpha\beta}(Z_j\tl{Z}_j^\dagger)^\alpha(X_j\tl{X}_j^\dagger)^\beta.\label{sw2}
\end{eqnarray}
whose action on a local spin reads
\begin{eqnarray}
    S_jX_j=\tl{X}_j,\quad  S_jZ_j=\tl{Z}_j,\quad  S_j\tl{X}_j=X_j,\quad  S_j\tl{Z}_j=Z_j.
\end{eqnarray}
It is straightforward to verify that the Hamiltonian~\eqref{eq:spin} commutes with an operator
$S\vcentcolon=\prod_{j=1}^LS_j$. 
Investigation given below in this subsection remains valid as long as the model respects the symmetry~\eqref{dip} and $[H,S]=0$. One could add interaction terms, such as 
\begin{eqnarray*}
    -g\sum_jZ_{j-1}(Z_j^\dagger)^2Z_{j+1}\times \tl{Z}_{j-1}(\tl{Z}_j^\dagger)^2\tl{Z}_{j+1}
    +h.c.
\end{eqnarray*}
which is the dipole analog of the interaction found in the Ashkin-Teller model,
to the Hamiltonian~\eqref{dip}. 
\par
Now we are in a good place to perform gauging the $0$-form dipole symmetry $\tl{Q}_0$ and $\tl{Q}_{dipole}$, viz $0$-form dipole symmetry for spins~$\tl{Z}_i$ and construct the operator.
To do so, we introduce an extended Hilbert space located at each node of the chain whose $\mathbb{Z}_N$~Pauli operator is denoted as $\tl{\tau}^X_j$ and $\tl{\tau}^Z_j$, corresponding to gauge fields. 
The Gauss law is given by 
\begin{eqnarray}
    (\tl{\tau}^X_{j-1})^\dagger\tl{\tau}^X_j \tl{X}_j \tl{\tau}^X_j(\tl{\tau}^X_{j+1})^\dagger  =1.\label{gauss dip}
\end{eqnarray}
Intuitive understanding of the Gauss law term~\eqref{gauss dip} is that one decomposes the global charges~\eqref{dip} into local ones by introducing extended Hilbert spaces associated with the gauge fields. To see how, we define 
\begin{eqnarray*}
    \mathcal{G}_{1Ddip,j}\vcentcolon= (\tl{\tau}^X_{j-1})^\dagger\tl{\tau}^X_j \tl{X}_j \tl{\tau}^X_j(\tl{\tau}^X_{j+1})^\dagger  
\end{eqnarray*}
and obtain the following:
\begin{eqnarray}
    \tl{Q}_0=\prod_{j=1}^{L}\mathcal{G}_{1Ddip,j},\quad \tl{Q}_{dipole}=\prod_{j=1}^{L}(\mathcal{G}_{1Ddip,j})^j.
\end{eqnarray}
The Gauss law is imposed by setting $\mathcal{G}_{1Ddip,j}=1\quad\forall j$.\par
In order for other terms to commute with the Gauss law~\eqref{gauss dip}, we minimally couple the matter terms to the gauge field as
\begin{eqnarray}
    \tl{Z}_{j-1}(\tl{Z}_j^\dagger)^2\tl{Z}_{j+1} \to   \tl{Z}_{j-1}(\tl{Z}_j^\dagger)\tau^Z_j(\tl{Z}_j^\dagger)\tl{Z}_{j+1}
\end{eqnarray}
To proceed, we define new variables as
\begin{eqnarray*}
    \tau^X_{j}\vcentcolon   = \tl{\tau}^X_{j},\quad  \tau^Z_{j}\vcentcolon =  \tl{Z}_{j-1}(\tl{Z}_j^\dagger)\tau^Z_j(\tl{Z}_j^\dagger)\tl{Z}_{j+1}.
\end{eqnarray*}
After gauging, 
we obtain the following map:
\begin{eqnarray}
    \tl{X}_{j}\Rightarrow   G_j,\quad  \tl{Z}_{j-1}(\tl{Z}_j^\dagger)^2\tl{Z}_{j+1}\Rightarrow \tau^Z_j,\label{map0}
\end{eqnarray}
where 
\begin{eqnarray}
    G_j\vcentcolon =\tau^X_{j-1}(\tau^{X\dagger}_j)^2\tau^X_{j+1}.
\end{eqnarray}
The gauged Hamiltonian becomes
\begin{eqnarray}
     \widehat{H}_{1D:dipole}=-J\sum_j Z_{j-1}(Z_j^\dagger)^2Z_{j+1}-h\sum_jX_j-J\sum_{j}\tz_j-h\sum_{j}G_j
    +h.c.\label{gauged24}
\end{eqnarray}
Note that the gauged Hamiltonian~\eqref{gauged24} respects the following emergent $0$-form dipole symmetries:\footnote{These are the emergent quantum symmetries after gauging global symmetries.}
\begin{eqnarray}
    \eta_0=\prod_{j=1}^L\tz_j,\quad  \eta_{dipole}=\prod_{j=1}^L(\tz_j)^j\label{62}
\end{eqnarray}
with the same relation as~\eqref{64} where we replace $Q_0$ and $Q_{dipole}$ with $\eta_0$ and $\eta_{dipole}$, respectively. 
\par
Similar to the discussion presented in Sec.~\ref{section2}, one would expect that the operator $S$ plays the role of the operator, sending $X_j$ and $Z_{j-1}(Z_j^\dagger)^2Z_{j+1}$ to $G_j$ and $\tau^Z_j$, respectively, based on the 
fact that $S$ transforms a spin operator without tilde into the one with tilde before the gauging and these are further mapped to the operators according to~\eqref{map0} after gauging. To see this is correct, we need to carefully check the form of $S$ after gauging.
After gauging the $0$-form dipole symmetry, 
 a local spin flip term $\tl{X}_j$ in~\eqref{sw2} is replaced by $G_j$ due to the Gauss law~\eqref{gauss dip}.
However, since $\tl{Z}_j$ does not commute with the $0$-form dipole symmetry~\eqref{dip}, it becomes non-local after gauging, involving multiplication of the gauge fields. 
To resolve this issue, one replace a local spin $\tl{Z}_j$ with the Wilson operator $W_{j}$ described by nonlocal string of the gauge fields, meaning, 
 \begin{eqnarray}
    W_j= \prod _{0\leq k\leq j}\left(\tau^Z_k\right)^{j}\prod_{0\leq k\leq j}\left(\tau^Z_k\right)^{-k}.
 \end{eqnarray}
The first product represents a homogeneous string of the gauge fields whereas the second one does spatially modulated string of the gauge fields, depending on the coordinate of the spin.\footnote{Such an inhomogeneous string of the gauge fields reminds us of the Wilson line of dipole of anyons whose intensity depends linearly on the coordinate discussed in lattice spin models and BF theories with dipole symmetries~\cite{ebisu2209anisotropic,2023foliated}. 
 For clarity,~$k$ is a dummy index whereas $j$ denotes a coordinate of the spin.}
 The swap operator~\eqref{sw2} now becomes
 \begin{eqnarray}
     S_j=\frac{1}{N}\sum_{\alpha,\beta=1}^N\omega^{-\alpha\beta}\left[Z_j\prod _{0\leq k\leq j}\left(\tau^Z_k\right)^{-j}\prod_{0\leq k\leq j}\left(\tau^Z_k\right)^{k}
     \right]^\alpha\times\left[X_jG_j^\dagger\right]^\beta.
 \end{eqnarray}
 Note that operators $\{S_j|1\leq j\leq L-2\}$ commute with one another whereas $S_{L-1}$ does not with $S_{j}(j\neq 0\mod N)$. Also, $S_L$ does not commute with $S_j(j\neq 1\mod N)$.\par
We define the total swap operator as
 \begin{eqnarray}
     S\vcentcolon=S_1\times S_2\times\cdots\times S_{L-1} \times S_L,\label{pp}
 \end{eqnarray}
We check whether the operator $S$ after gauging is the operator.
After some algebra, one finds 
 \begin{eqnarray}
     SX_j=G_jS,\quad SG_j=X_jS,\quad S\tau^Z_j=Z_{j-1}(Z_j^\dagger )^2 Z_{j+1}S,\quad S Z_{j-1} (Z_j^\dagger)^2 Z_{j+1}=\tau^Z_j S\quad (2\leq j\leq L-1),
 \end{eqnarray}
 hence, away from the point $j=1,L$, the product of the swap operator is the desired operator. However, 
 taking care of the periodic boundary condition, we have
 \begin{align}
&SX_L=G_LS{Q}_0{Q}_{dipole}^\dagger,\quad
     SX_1=G_1S{Q}_{dipole}^\dagger,
     \quad&& SG_L=X_L S,\quad SG_1=X_1S, \nonumber
     \\
     &SZ_{L-1}(Z_L^\dagger)^2 Z_1=\tz_LS\eta_0\eta_{dipole},\quad&& 
     SZ_{L}(Z_1^\dagger)^2 Z_2=\tz_1S\eta_{dipole}^\dagger,\\
    &S\tz_L= Z_1 (Z_L^\dagger)^2 Z_{L-1}S\eta_0\eta_{dipole}^\dagger,\quad&&  S\tz_1=  Z_L (Z_1^\dagger)^2 Z_{2} S\eta_{dipole}. \nonumber
 \end{align}
implying that when acting the product of the swap operators on spins, they are transformed into the ones corresponding to the mapping under the duality, accompanied by global charges, ${Q}_0$, ${Q}_{dipole}$, $\eta_0$, and~$\eta_{dipole}$. To fix this issue, we introduce the following operator:
\begin{eqnarray}
    D\vcentcolon=\frac{1}{N}S\times\left[\sum_{\alpha,\beta,\gamma,\delta=1}^N {Q}^\alpha_0{Q}^\beta_{dipole}\eta_0^\gamma\eta_{dipole}^\delta\right].
\end{eqnarray}
This operator is the desired swap operator, which can be verified by
\begin{eqnarray}
      DX_j=G_jD,\quad DG_j=X_jD,\quad D\tau^Z_j= Z_{j-1} (Z_j^\dagger)^2 Z_{j+1} D,\quad D  Z_{j-1} (Z_j^\dagger)^2 Z_{j+1}=\tau^Z_j D\quad (1\leq j\leq L).
\end{eqnarray}
Further, the operator is noninvertible; we obtain the following fusion rules: 
\begin{align}
 D\times D=\sum_{\alpha,\beta,\gamma,\delta=1}^N {Q}^\alpha_0{Q}^\beta_{dipole}\eta_0^\gamma\eta_{dipole}^\delta,\quad DP=PD=D\;(P={Q}_0,{Q}_{dipole},\eta_0,\eta_{dipole}),
 \label{77}
\end{align}
implying the fusion rule of the operators involves dipole and global charges. Note that previous study showed that fusion rule of the noninvertible operator obtained by gauging $0$-form dipole symmetry in a single chain yields charge conjugation operator in addition to the dipole and global charges.~\cite{Cao:2024qjj,Pace:2024tgk}. 
As opposed to the previous considerations, in our setting comprised of double copies of the spin chains, there is no charge conjugation operator in the fusion rule. 
\subsection{Two dimensions}
In this subsection, we demonstrate construction of noninvertible operators in two dimensions via gauging $0$-form dipole symmetry. 
A new feature of such operators compared with the case of the one dimension is that we have variety of hierarchical structures of dipole and global charges that we dub \textit{dipole algebra}.  Note that the procedure of gauging such symmetries in two dimensions were studied in a field theoretical approach~\cite{2023foliated,anomaly_2024}, yet, to our knowledge, the explicit procedure to perform gauging  these symmetries at the lattice level has not been discussed previously.
\subsubsection{Model}
\begin{figure}
    \begin{center}
       
       \begin{subfigure}[h]{0.49\textwidth}
       \centering
  \includegraphics[width=0.9\textwidth]{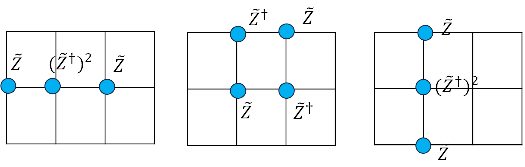}
         \caption{}\label{disk}
             \end{subfigure}
            \begin{subfigure}[h]{0.49\textwidth}
            \centering
  \includegraphics[width=1\textwidth]{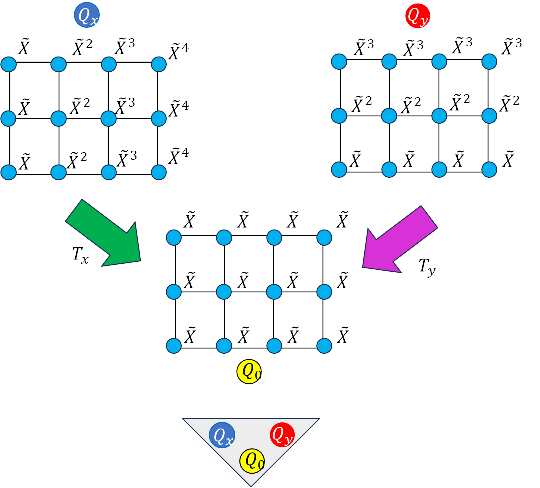}
         \caption{}\label{disk3}
             \end{subfigure}
                          \hspace{29mm}

                    \begin{subfigure}[h]{0.49\textwidth}
                    \centering
  \includegraphics[width=0.7\textwidth]{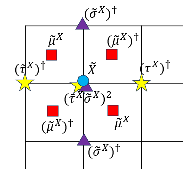}
         \caption{}\label{disk2}
             \end{subfigure}
                               \begin{subfigure}[h]{0.49\textwidth}
                               \centering
  \includegraphics[width=0.7\textwidth]{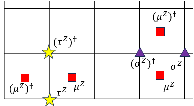}
         \caption{}\label{disk4}
             \end{subfigure}
 \end{center}
 \caption{(a)~Three types of terms defined in~\eqref{spin2d2}. 
 (b) $0$-form dipole symmetries~\eqref{algebra}, forming dipole algebra~\eqref{algebra2} which is schematically portrayed as an ``inverse of a triangle'' in the bottom.
 (c)~The Gauss law corresponding to~\eqref{gauss_44}. (d) Two flux operators given in~\eqref{fluxes}. }\label{fig4}
 \end{figure}
 \par
 To set the stage, we envisage a 2D lattice model and introduce two $\mathbb{Z}_N$ spins on each node. 
 For simplicity, we set the system size of the lattice as $L_x\times L_y$ with 
 $L_x\equiv L_y\equiv 0\mod N$ and impose the periodic boundary condition.
 We consider the following Hamiltonian:
 \begin{eqnarray}
     H_{2D:dipole}=&-&J_x\sum_{\br}\mathcal{N}^{Z}_{x,\br}-J_y\sum_{\br}\mathcal{N}^Z_{y,\br}-J_{xy}\sum_{\bp}\mathcal{P}^Z_{\bp}-h\sum_{\br}X_{\br}\nonumber\\
     &-&\tl{J}_x\sum_{\br}\tl{\mathcal{N}}^{Z}_{x,\br}-\tl{J}_y\sum_{\br}\tl{\mathcal{N}}^Z_{y,\br}-\tl{J}_{xy}\sum_{\bp}\tl{\mathcal{P}}^Z_{\bp}-\tl{h}\sum_{\br}\tl{X}_{\br}
    +h.c.,\label{spin2d}
 \end{eqnarray}
 with
 \begin{equation}
 \begin{split}
    \mathcal{N}^Z_{x,\br}\vcentcolon=& Z_{\br-\exx}(Z^\dagger_{\br})^2Z_{\br+\exx},\quad    \mathcal{N}^Z_{y,\br}\vcentcolon=Z_{\br-\eyy}(Z^\dagger_{\br})^2Z_{\br+\eyy},\\
   \mathcal{P}^Z_{\bp}\vcentcolon=& Z^\dagger_{\bp-\ex-\ey}Z_{\bp+\ex-\ey}Z^\dagger_{\bp+\ex+\ey}Z_{\bp-\ex+\ey}\label{spin2d2}
   \end{split}
 \end{equation}
 and similarly for $\tl{\mathcal{N}}^Z_{x,\br}$, $\tl{\mathcal{N}}^Z_{y,\br}$, and $\tl{\mathcal{P}}^Z_{\bp}$ obtained by replacing $Z_{\br}$ with $\tl{Z}_{\br}$. 
The terms in~\eqref{spin2d2} are depicted in Fig.~\ref{disk}. 
In what follows, we concentrate on a situation where $J_x=\tl{J}_x$, $J_y=\tl{J}_y$, $J_{xy}=\tl{J}_{xy}$, and $h=\tl{h}$. In such a case, 
the Hamiltonian~\eqref{spin2d} respects the following $0$-form dipole symmetries~(see also Fig.~\ref{disk3}):
\begin{eqnarray}
    Q_{2D:0}=\prod_{\hx=1}^{L_x}\prod_{\hy=1}^{L_y}X_{\br},\quad Q_{2D:x}=\prod_{\hx=1}^{L_x}\prod_{\hy=1}^{L_y}(X_{\br})^{\hx},\quad Q_{2D:y}=\prod_{\hx=1}^{L_x}\prod_{\hy=1}^{L_y}(X_{\br})^{\hy}\label{algebra}
\end{eqnarray}
with relation 
\begin{align}
    T_xQ_{2D:x}T_x^{-1}&=Q_{2D:x}Q_{2D:0}^\dagger,\quad
    T_yQ_{2D:y}T_y^{-1}=Q_{2D:y}Q_{2D:0}^\dagger, \nonumber\\
    T_xQ_{2D:0}T_x^{-1}&= T_yQ_{2D:0}T_y^{-1}=Q_{2D:0},\quad  T_xQ_{2D:y}T_x^{-1}= Q_{2D:y},\quad T_yQ_{2D:x}T_y^{-1}= Q_{2D:x},
    \label{algebra2}
\end{align}
where $T_{x}$ ($T_y$) denotes lattice translational operator in the $x$($y$)-direction. To be more precise, 
$T_xX_{\br}T_x^{-1}=X_{\br+\exx}$, $T_yX_{\br}T_y^{-1}=X_{\br+\eyy}$.
The charges~\eqref{algebra} with relation~\eqref{algebra2} correspond to 
the lattice analog of the dipole algebra in a field theory~\cite{2023foliated}, consisting of two dipole charges followed by a global one, forming a hierarchy; 
a global charge~$Q_{2D:0}$ is generated by acting the translational operator in the $x(y)$-direction on the dipole charge $Q_{2D:x}~(Q_{2D:y})$. Such a hierarchy
is symbolically represented by an ``inverse of a triangle'' given in the bottom of Fig.~\ref{disk3} with interpretation that what sits at the bottom is generated by the one located above via translational operator. The model~\eqref{spin2d} admits the similar $0$-form dipole symmetry described by charges $\tl{Q}_{2D:0}$, $\tl{Q}_{2D:x}$, and $\tl{Q}_{2D:y}$ which are obtained by replacing $X_{\br}$ with $\tl{X}_{\br}$ in~\eqref{algebra}. Further, the Hamiltonian~\eqref{spin2d} respects $\mathbb{Z}_2$ symmetry, exchanging $X_{\br}$ and $\tl{X}_{\br}$ as well as $Z_{\br}$ and $\tl{Z}_{\br}$. To wit, the Hamiltonian~\eqref{spin2d}
commutes with $S=\prod_{\br}S_{\br}$, where
\begin{eqnarray}
    S_{\br}\vcentcolon=\frac{1}{N}\sum_{\alpha,\beta=1}^N\omega^{-\alpha\beta}(Z_{\br}\tl{Z}_{\br}^\dagger)^\alpha(X_{\br}\tl{X}_{\br}^\dagger)^\beta.\label{sbr}
\end{eqnarray}
The following discussion remains valid as long as the model respects this $\mathbb{Z}_2$ symmetry in addition to the $0$-form dipole symmetries~\eqref{algebra} for two spin degrees of freedom. For instance, one could add interaction terms to preserve these symmetries.
\par
Now we gauge $0$-form dipole symmetry, especially the ones for the spins $\tl{Z}_{\br}$. To do so, we introduce three types of the Hilbert space, two of which are located on each node and the other is on each plaquette. Denoting Pauli operators, corresponding to these Hilbert spaces as $\tl{\tau}^{X/Z}_{\br}$, $\tl{\sigma}^{X/Z}_{\br}$, $\tl{\mu}^{X/Z}_{\bp}$, the Gauss law is given by (see also Fig.~\ref{disk2})
\begin{eqnarray}
    \tl{\tau}^{X\dagger}_{\br-\exx}(\tl{\tau}^{X}_{\br})^2 \tl{\tau}^{X\dagger}_{\br+\exx}\times \tl{\sigma}^{X\dagger}_{\br-\eyy}(\tl{\sigma}^{X}_{\br})^2 \tl{\sigma}^{X\dagger}_{\br+\eyy}\times \tl{\mu}^X_{\bp-\ex+\ey}(\tl{\mu}^X_{\bp-\ex-\ey})^{\dagger}\tl{\mu}^X_{\bp+\ex-\ey}(\tl{\mu}^X_{\bp+\ex+\ey})^{\dagger}\tl{X}_{\br}=1.\label{gauss_44}
\end{eqnarray}
The intuition behind the Gauss law term~\eqref{gauss_44} is that one decomposes charges~\eqref{algebra} into local one with introducing extended Hilbert spaces, corresponding to gauge fields. This is in line with the fact that gauging is a procedure to promote global symmetries into local ones, and practically, on a lattice one "crops" a global spin flip into a local spin flip with inclusion of extended Hilbert spaces.
Introducing
\begin{eqnarray}
    \mathcal{G}_{2Ddip,\br}\vcentcolon=   \tl{\tau}^{X\dagger}_{\br-\exx}(\tl{\tau}^{X}_{\br})^2 \tl{\tau}^{X\dagger}_{\br+\exx}\times \tl{\sigma}^{X\dagger}_{\br-\eyy}(\tl{\sigma}^{X}_{\br})^2 \tl{\sigma}^{X\dagger}_{\br+\eyy}\times \tl{\mu}^X_{\bp-\ex+\ey}(\tl{\mu}^X_{\bp-\ex-\ey})^{\dagger}\tl{\mu}^X_{\bp+\ex-\ey}(\tl{\mu}^X_{\bp+\ex+\ey})^{\dagger}\tl{X}_{\br},
\end{eqnarray}
one finds
 \begin{eqnarray}
Q_{2D:0}=\prod_{\hx=1}^{L_x}\prod_{\hy=1}^{L_y}\mathcal{G}_{2Ddip,\br},\quad Q_{2D:x}=\prod_{\hx=1}^{L_x}\prod_{\hy=1}^{L_y}(\mathcal{G}_{2Ddip,\br})^{\hx},\quad Q_{2D:y}=\prod_{\hx=1}^{L_x}\prod_{\hy=1}^{L_y}(\mathcal{G}_{2Ddip,\br})^{\hy}.
\end{eqnarray}
The Gauss law is imposed by setting $\mathcal{G}_{2Ddip,\br}=1\quad\forall\br$.
\par
We minimally couple the matter terms to the gauged fields via
\begin{eqnarray}
   \tl{\mathcal{N}}^Z_{x,\br}\to  \tl{\mathcal{N}}^Z_{x,\br}\tl{\tz}_{\br},\quad  \tl{\mathcal{N}}^Z_{y,\br}\to  \tl{\mathcal{N}}^Z_{y,\br}\tl{\sigma}^Z_{\br},\quad
    \tl{\mathcal{P}}^Z_{\bp}\to  \tl{\mathcal{P}}^Z_{\bp}\tl{\mu}^Z_{\bp}.
\end{eqnarray}
Defining new variables as
\begin{align}
    \tau^X_{\br}\vcentcolon=&\tl{\tau}^X_{\br},&&  \sigma^X_{\br}\vcentcolon=\tl{\sigma}^X_{\br},&&\mu^X_{\bp}\vcentcolon=\tl{\mu}^X_{\bp} ,\nonumber\\
       \tau^Z_{\br}\vcentcolon=&\tl{\mathcal{N}}^Z_{x,\br}\tl{\tz}_{\br},&&\sigma^Z_{\br}\vcentcolon=\tl{\mathcal{N}}^Z_{y,\br}\tl{\sigma}^Z_{\br},&&\mu^Z_{\bp}\vcentcolon=\tl{\mathcal{P}}^Z_{\bp}\tl{\mu}^Z_{\bp},
\end{align}
we have the following mapping via gauging:
\begin{eqnarray}
    \tl{X}_{\br}\Rightarrow G_{\br}^X,\quad  \tl{\mathcal{N}}^Z_{x,\br}\Rightarrow \tau^Z_{\br},\quad\tl{\mathcal{N}}^Z_{y,\br}\Rightarrow \sigma^Z_{\br},\quad  \tl{\mathcal{P}}^Z_{\bp}\Rightarrow \mu^Z_{\bp},\label{mapping}
\end{eqnarray}
where 
\begin{eqnarray}
    G^X_{\br}\vcentcolon={\tx}_{\br-\exx}({\tau}^{X\dagger}_{\br})^2{\tx}_{\br+\exx}\times {\sigma}^X_{\br-\eyy}({\sigma}^{X\dagger}_{\br})^2 {\sigma}^X_{\br+\eyy}\times ({\mu}^{X}_{\bp-\ex+\ey})^\dagger{\mu}^{X}_{\bp-\ex-\ey}({\mu}^X_{\bp+\ex-\ey})^\dagger{\mu}^X_{\bp+\ex+\ey}.\label{gsss}
\end{eqnarray}
Further, we add the following flux operators to~\eqref{spin2d} to
make sure that gauge theory is dynamically trivial (viz flatness condition of the gauge fields is imposed):
\begin{eqnarray}
&-& g_{B_x}\sum_{\bx}B_{\bx}-g_{B_y}\sum_{\by}B_{\by}
 +h.c.\vcentcolon= \nonumber \\
 &-& g_{B_x}\sum_{\bx}\sigma^{Z\dagger}_{\bx-\ex}\sigma^Z_{\bx+\ex}\mu^Z_{\bx-\ey}\mu^{Z\dagger}_{\bx+\ey}
-g_{B_y}\sum_{\by}\mu^{Z\dagger}_{\by-\ex}\mu^Z_{\by+\ex}\tau^Z_{\by-\ey}\tau^{Z\dagger}_{\by+\ey}+h.c.
 \label{fluxes}
\end{eqnarray}
Here, $\bx\vcentcolon=(\hx+\frac{1}{2},\hy)$, and $\by\vcentcolon=(\hx,\hy+\frac{1}{2})$. Overall, the gauged Hamiltonian reads
\begin{eqnarray}
       \widehat{H}_{2D:dipole}=\widehat{H}_0+\widehat{H}_{DTC}\label{2d}
\end{eqnarray}
with
\begin{eqnarray}
    \widehat{H}_0\vcentcolon=&-&J_x\sum_{\br}\mathcal{N}^{Z}_{x,\br}-J_y\sum_{\br}\mathcal{N}^Z_{y,\br}-J_{xy}\sum_{\bp}\mathcal{P}^Z_{\bp}\nonumber\\
    &-&h\sum_{\br}X_{\br}-{J}_x\sum_{\br}\tau^Z_{\br}-{J}_y\sum_{\br}\sigma^Z_{\br}-
     J_{xy}\sum_{\bp}\mu^Z_{\bp}
    +h.c.,\label{88}\\
    \widehat{H}_{DTC}\vcentcolon=&-&{h}\sum_{\br}G^X_{\br} - g_{B_x}\sum_{\bx}B_{\bx}-g_{B_y}\sum_{\by}B_{\by}+h.c.\label{89}
\end{eqnarray}
After gauging $0$-form dipole symmetry, one obtains the $\mathbb{Z}_N$ toric code with dipole symmetry, $\widehat{H}_{DTC}$,~\cite{PhysRevB.97.235112,PhysRevB.106.045145,oh2022rank}, corresponding to the fact that the usual $\mathbb{Z}_N$ toric code~\cite{KITAEV20032} is obtained by gauging ordinary global symmetry.
\subsubsection{Toric code with dipole symmetries}
For later convenience, we briefly recall the properties of the model,~$\widehat{H}_{DTC}$,~\eqref{89} 
in the limit of~$h, g_{B_x}, g_{B_y}\to\infty$, 
before delving into the construction of the noninvertible operator. \par
Similar to the standard toric code, the model~\eqref{89} is the exactly solvable as each terms in the Hamiltonian commutes with one another. Also, the ground state is the projected state~$\ket{\omega}$, satisfying, 
\begin{equation}
G^X_{\br}\ket{\omega}=B_{\bx}\ket{\omega}=B_{\by}\ket{\omega}=\ket{\omega}\quad\forall~\br,\bx,\by.\label{omega}
\end{equation}
On the torus geometry with system size $L_x\times L_y$ ($L_x\equiv L_y\equiv 0\mod N$), the model exhibits the nontrivial ground state degeneracy (GSD). To see how, we count the distinct number of noncontractible loops of the operators $\tau^Z_{\br}$, $\sigma^Z_{\br}$, and $\mu^Z_{\bp}$. There are three types noncontractible loops in the $x$-direction, such loops are given by
\begin{eqnarray}
    \xi^Z_{1}\vcentcolon=\prod_{\hx=1}^{L_x}\tau^Z_{(\hx,0)},\quad \xi^Z_2\vcentcolon=\prod_{\hx=1}^{L_x}\left(\tau^{Z\dagger}_{(\hx,1)}\right)^{\hx}\left(\tau^{Z}_{(\hx,0)}\right)^{\hx},\quad \xi^Z_{x,y}\vcentcolon=\prod_{\hx=1}^{L_x}\mu^Z_{(\hx+\frac{1}{2},\frac{1}{2})}\left(\tau^{Z}_{(\hx,0)}\right)^{\hx}.\label{loops}
\end{eqnarray}
These loops are shown in Fig.~\ref{disk5}. A simple calculation, jointly with~\eqref{omega} leads to that 
\begin{align}
T_x\xi^Z_{x,y}T_x^{-1}&=\xi^{Z\dagger}_1\xi^Z_{x,y},\quad&&  T_y\xi^Z_{x,y}T_y^{-1}=\xi^{Z\dagger}_2\xi^Z_{x,y},\nonumber\\
    T_x\xi^Z_iT_x^{-1}&=\xi^Z_i,\quad&& T_y\xi^Z_iT_y^{-1}=\xi^Z_i~(i=1,2).\label{dual94}
\end{align}
This indicates that 
the loops~\eqref{loops} constitute $1$-form \textit{dual dipole algebra}~\cite{anomaly_2024}, that is, one $1$-form dipole charge followed by two $1$-form charges, symbolically described by a triangle portrayed in the bottom of Fig.~\ref{disk5}. 
\par
It is known that when gauging a $p$-form symmetry in $d$ spatial 
dimensions, one has an emergent $(d-p-1)$-form dual symmetry~\cite{gaiotto2015generalized}. As an example in the context of the condensed matter physics, when gauging the ordinary $0$-form $\mathbb{Z}_2$ symmetry in the 2D paramagnetic Ising model, one obtains the $\mathbb{Z}_2$ toric code, which has an emergent $1$-form dual symmetry  associated with noncontractible loops of 
fractional
quasiparticle excitations. Here, we have seen one example that extends this fact to the dipole symmetry; gauging $0$-form dipole symmetry with a dipole algebra gives rise to $1$-form dual dipole symmetry where hierarchical structure of dipole algebra is inverted~(See Fig.~\ref{dl}).~\footnote{Such a phenomenon is also consistent with the result in one dimension~(Sec.~\ref{subsec:1ddipole}). Indeed, gauging $0$-form dipole symmetries $\tilde{Q}_0$ and $\tilde{Q}_{dipole}$ in Eq.~\eqref{dip} yields emergent $0$-form dual dipole symmetries $\eta_{dipole}$ and $\eta_0$ in Eq.~\eqref{62}, respectively. 
This can be interpreted as that via gauging $0$-form dipole symmetry with a dipole algebra~\eqref{dip}, one has the emergent $0$-form dipole symmetry with the dual dipole algebra with the hierarchical structure is inverted, that is, the role of dipole and global charges are switched, giving~\eqref{62}.
}\par
One also finds that there are three types of noncontractible loops of the gauge fields that go along the~$y$-direction:
   \begin{eqnarray}
    \eta^Z_{1}\vcentcolon=\prod_{\hy=1}^{L_y}\sigma^Z_{(0,\hy)},\quad \eta^Z_2\vcentcolon=\prod_{\hy=1}^{L_y}\left(\sigma^{Z\dagger}_{(1,\hy)}\right)^{\hy}\left(\sigma^{Z}_{(0,\hy)}\right)^{\hy},\quad \eta^Z_{x,y}\vcentcolon=\prod_{\hy=1}^{L_y}\mu^Z_{(\frac{1}{2},\hy+\frac{1}{2})}\left(\sigma^{Z}_{(0,\hy)}\right)^{\hy}.\label{loopy}
\end{eqnarray}
with the following relations
\begin{align}
    T_x\eta^Z_{x,y}T_x^{-1}&=\eta^{Z\dagger}_2\eta^Z_{x,y},\quad&&  T_y\eta^Z_{x,y}T_y^{-1}=\eta^{Z\dagger}_1\eta^Z_{x,y},\nonumber\\
    T_x\eta^Z_iT_x^{-1}&=\eta^Z_i,\quad&& T_y\eta^Z_iT_y^{-1}=\eta^Z_i~ \qquad (i=1,2),\label{dual942}
\end{align}
implying that these loops also form $1$-form dipole symmetry with dual dipole algebra. 
Six distinct noncontractible loops defined in~\eqref{loops} and~\eqref{loopy} contribute to the nontrivial GSD, which is given by $N^6$. We would like to emphasize that although the GSD of the toric code with dipole symmetry was identified in UV lattice and field theoretical models~\cite{PhysRevB.97.235112,PhysRevB.106.045145,oh2022rank,2023foliated}, the discussion on the model in view of dipole algebra, especially, $1$-form dual dipole algebra~\eqref{dual94} has not been given previously. Such a perspective will be crucial in understanding operators that we will turn to next.

\subsubsection{noninvertible operators}\label{423}
\begin{figure}
                 \begin{subfigure}[h]{0.49\textwidth}
  \includegraphics[width=1\textwidth]{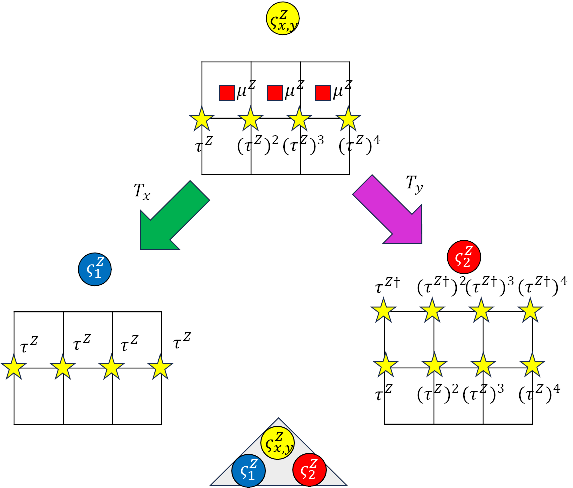}
         \caption{}\label{disk5}
             \end{subfigure}
               \begin{subfigure}[h]{0.49\textwidth}
  \includegraphics[width=1\textwidth]{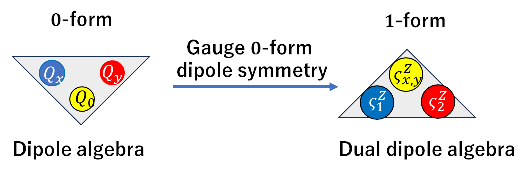}
         \caption{}\label{dl}
             \end{subfigure}
 \begin{subfigure}[h]{1\textwidth}
  \centering
  \includegraphics[width=0.2\textwidth]{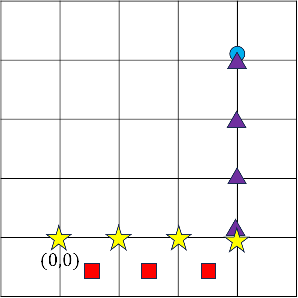}
         \caption{}\label{st}
             \end{subfigure}
          \caption{
          (a)~Configuration of loops given in~\eqref{loops}. These loops constitute $1$-form dual dipole algebra. (b)~
          When gauging $0$-form dipole symmetry characterized by a dipole algebra~(left) yields $1$-form dipole symmetry labeled by dual dipole algebra~(right). 
          (c)~Configuration of gauge fields defined in~\eqref{95}.}\label{fig5}
              \end{figure}

After review the properties of the toric code with dipole symmetry~\eqref{89}, we construct the operators in the model~\eqref{2d}. The way we obtain these closely parallels the ones in the previous sections. 
After gauging the $0$-form dipole symmetry, we replace a local operator $\tl{Z}_{\br}$ with the Wilson operator~$\mathcal{T}_{\br}$ which has the following form:
\begin{eqnarray}
\mathcal{T}_{\br}=\underbrace{\left[\prod_{0\leq \hx^\prime\leq \hx}\left(\tau^Z_{(\hx^\prime,0)}\right)^{\hx}\left(\tau^Z_{(\hx^\prime,0)}\right)^{-\hx^\prime}\right]}_{\bigstar}\times\underbrace{\left[\prod_{0\leq \hy^\prime\leq \hy}\left(\sigma^Z_{(\hx,\hy^\prime)}\right)^{\hy}\left(\sigma^Z_{(\hx,\hy^\prime)}\right)^{-\hy^\prime}\right]}_{\triangle}\times\underbrace{\prod_{0\leq \hx^\prime\leq \hx}\left(\mu^{Z\dagger}_{(\frac{1}{2}+\hx^\prime,-\frac{1}{2})}\right)^{\hy}}_{\square}.\label{95}
\end{eqnarray}
The right hand side of~\eqref{95} consists of string of the gauge fields, $\tau^Z_{\br}$, $\sigma^Z_{\br}$, and $\mu^Z_{\bp}$ each of 
which is depicted by $\bigstar$, $\triangle$, and $\square$  in Fig.~\ref{st}. Accordingly, the operator $S_{\br}$ given in~\eqref{sbr} becomes
\begin{eqnarray}
    S_{\br}=\frac{1}{N}\sum_{\alpha,\beta=1}^N\omega^{-\alpha\beta}\left[Z_{\br}\mathcal{T}_{\br}^\dagger
     \right]^\alpha\times\left[X_{\br}G_{\br}^{X\dagger}\right]^\beta.
\end{eqnarray}
One can verify that the operators $S_{\br}$~$(\br\neq (0,0),(L_x-1,0),(0,L_y-1))$ commute with one another yet do not with the ones on the boundary, that is, $S_{(0,0)}$, $S_{(L_x-1,0)}$, and $S_{(0,L_y-1)}$. 
We define the product of $S_{\br}$ as
\begin{eqnarray}
      S=\prod_{\br\neq (0,0),(L_x-1,0),(0,L_y-1) }S_{\br}\times \left(S_{(0,0)}\times
   S_{(L_x-1,0)}\times S_{(0,L_y-1)}
    \right).
\end{eqnarray}
Note that the order of the product is important due to the commutation relations mentioned above. One finds that the action of $S$ on local spins, $X_{\br}$, $G^X_{\br}$ as follows:
\begin{eqnarray}
    SX_{\br}&=&G^X_{\br}S,\quad[\br\neq (0,0),(L_x-1,0),(0,L_y-1)]\nonumber\\
    SG^X_{\br}&=&X_{\br}S\quad\forall\br.
\end{eqnarray}
However, the action of $S$ on the spins $X_{\br}$ around the boundary picks up global dipole charges. Indeed, one can verify that 
\begin{align}
    SX_{(0,0)}&=G^X_{(0,0)}SQ_{2D:0}^\dagger Q_{2D:x}^\dagger Q_{2D:y},\quad SX_{(L_x-1,0)}=G^X_{(L_x-1,0)}SQ_{2D:x},\quad SX_{(0,L_y-1)}=G^X_{(0,L_y-1)}SQ_{2D:y}.\label{qqq}
\end{align}
To make the operator $S$ an appropriate one implementing the gauge mapping, we utilize a composite operator $SC_Q := S\times \left[\sum_{\alpha,\beta,\gamma=1}^N Q_{2D:0}^\alpha Q_{2D:x}^\beta Q_{2D:y}^\gamma\right]$, which acts on spins as 
\begin{eqnarray*}
  SC_QX_{(0,0)}=G^X_{(0,0)}SC_Q, \quad
   SC_QX_{(L_x-1,0)}=G^X_{(L_x-1,0)}SC_Q,\quad SC_QX_{(0,L_y-1)}=G^X_{(0,L_y-1)}SC_Q.\label{qqqp}
\end{eqnarray*}
To proceed, we investigate the action of $SC_Q$ on the coupling terms $\mathcal{N}_{x,\br}^Z$, $\mathcal{N}_{y,\br}^Z$, and $\mathcal{P}_{\bp}^Z$. For instance,  the action of $SC_Q$ on $\mathcal{N}_{x,\br}^Z$ reads
\begin{eqnarray}
   SC_Q\mathcal{N}_{x,\br}^Z=
   \tau^Z_{(\hx,0)}V_{\br}SC_Q              \label{ha}
\end{eqnarray}
with
\begin{eqnarray}
    V_{\br}\vcentcolon=\left[\prod_{0\leq \hy^\prime\leq \hy}\left(\sigma^Z_{(\hx-1,\hy^\prime)}(\sigma^{Z\dagger}_{(\hx,\hy^\prime)})^2\sigma^Z_{(\hx+1,\hy^\prime)}\right)^{\hy}\left(\sigma^{Z\dagger}_{(\hx-1,\hy^\prime)}(\sigma^{Z}_{(\hx,\hy^\prime)})^2\sigma^{Z\dagger}_{(\hx+1,\hy^\prime)}\right)^{\hy^\prime}\right] \times \left(\mu^{Z}_{(\hx-\frac{1}{2},-\frac{1}{2})}\mu^{Z\dagger}_{(\hx-\frac{1}{2},-\frac{1}{2})}\right)^{\hy}.\label{106}
\end{eqnarray}
In order for~\eqref{ha} to be the desired mapping~\eqref{mapping}, we multiply $SC_Q$ with the following operators:
\begin{eqnarray}
    C_{1}=\frac{1}{N^{2L_xL_y}}\sum_{\gamma}\eta(\gamma),\label{c1}
\end{eqnarray}
where $\sum_{\gamma}\eta(\gamma)$ represents sum over all kinds of loops of the gauge fields, $\tz_{\br}$, $\sigma^Z_{\br}$, $\mu^Z_{\bp}$, 
including the noncontractible ones. If we define an operator 
\begin{eqnarray}
    D\vcentcolon=\frac{1}{N}SC_QC_1, 
\end{eqnarray}
then one can verify that 
\begin{eqnarray}
    D \mathcal{N}_{x,\br}^Z=\tau^Z_{\br} D,
\end{eqnarray}
where we have used the fact that contractible loops of the gauge fields are generated by product of $B_{\bx}$'s and $B_{\by}$'s~\eqref{fluxes} and that multiplying some combination of $B_{\bx}$'s and $B_{\by}$'s with the operator $\tau^z_{(\hx,0)}V_{\br}$ gives~$\tau^Z_{\br}$.
While we carefully investigate whether $D$ acts on the coupling $\mathcal{N}_{x,\br}^Z$,  
the similar discussion holds for other coupling terms $\mathcal{N}_{y,\br}^Z$, $\mathcal{P}^Z_{\bp}$.
One can also show that (see Appendix.~\ref{app:1}) in the limit of $g_x,g_y\to\infty$, the operator $C_1$ becomes~\footnote{As a consistency check about the normalization prefactor $\frac{1}{N^3}$, we note the fact that the toric code with the dipole symmetry~\eqref{89} can be decomposed into three copies of the standard toric codes~\cite{2023foliated,2024multipole}, and that the normalization prefactor in the case of the single toric code gives $\frac{1}{N}$ as clarified in~\cite{Choi:2024rjm}. }
\begin{eqnarray}
    C_1=\frac{1}{N^3}\sum_{p,q,r=1}^N\sum_{s,t,u=1}^N(\xi_{x,y}^Z)^p (\xi_{1}^Z)^q(\xi_{2}^Z)^r
   ( \eta_{x,y}^Z)^s (\eta_{1}^Z)^t(\eta_{2}^Z)^u,\label{cond2}
\end{eqnarray}
where each $1$-form operator is given in~\eqref{loops} and~\eqref{loopy}. See Appendix.~\ref{app:1} for more details.
This is the dipole analog of the condensation defects~\cite{Roumpedakis:2022aik}. Note that three $1$-form operators, $\xi^Z_{x,y}, \xi^Z_{1},\xi^Z_{2}$ as well as $\eta^Z_{x,y}, \eta^Z_{1},\eta^Z_{2}$  form dual dipole algebra~\eqref{dual94}~\eqref{dual942}.
\par
To summarize, the operator $D$ is the desired swap operator, i.e., 
\begin{eqnarray}
   DX_{\br}=G^X_{\br}D,\quad D\mathcal{N}_{x,\br}^Z=\tau^Z_{\br} D,\quad  D\mathcal{N}_{y,\br}^Z=\sigma^Z_{\br} D,\quad  D\mathcal{P}_{\bp}^Z=\mu^Z_{\bp} D\nonumber\\
   DG^X_{\br}=X_{\br}D,\quad D\tau^Z_{\br}=\mathcal{N}_{x,\br}^ZD,\quad D\sigma^Z_{\br}=\mathcal{N}_{y,\br}^ZD,\quad D\mu^Z_{\bp}=\mathcal{P}_{\bp}^ZD,
\end{eqnarray}
and defining $C\vcentcolon=C_1C_Q$, we obtain the following
fusion rules
\begin{eqnarray}
    D\times D= C,\quad C\times C=N^6 C,\quad DC=CD=N^6D,\label{fusiondip}
\end{eqnarray}
indicating $D$ is the noninvertible operator. Further,~\eqref{fusiondip} is the dipole analog of 
fusion rules studied in the
$2$-category theory. The important distinction between the previous operators and the one in the present case is that when taking the limit $g_x,g_y\to\infty$, the operator $C$ is written as 
\begin{eqnarray}
    C=\left(\sum_{\alpha,\beta,\gamma=1}^N Q_{2D:0}^\alpha Q_{2D:x}^\beta Q_{2D:y}^\gamma\right)\times\left(\frac{1}{N^3}\sum_{p,q,r=1}^N\sum_{s,t,u=1}^N(\xi_{x,y}^Z)^p (\xi_{1}^Z)^q(\xi_{2}^Z)^r
   ( \eta_{x,y}^Z)^s (\eta_{1}^Z)^t(\eta_{2}^Z)^u\right),\label{fusiondip2}
\end{eqnarray}
indicating that $C$ consists of $0$-form global dipole charges with the dipole algebra and $1$-form dipole charges with the dual dipole algebra. Note that the hierarchical structure of the dipole and dual algebra is opposite to one another~(Fig.~\ref{dl}). 
\subsection{$0$-form dual dipole algebra in two dimensions}
\begin{figure}
    \begin{center}
       
       \begin{subfigure}[h]{0.49\textwidth}
       \centering
  \includegraphics[width=0.9\textwidth]{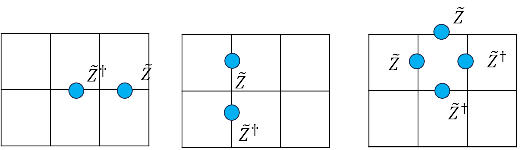}
         \caption{}\label{dualdipole}
             \end{subfigure}
            \begin{subfigure}[h]{0.49\textwidth}
            \centering
  \includegraphics[width=1.0\textwidth]{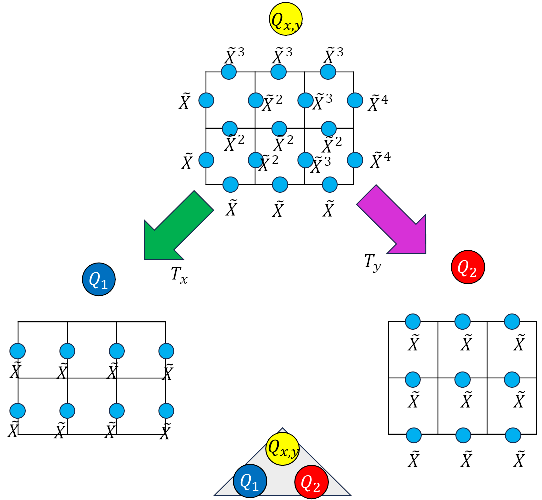}
         \caption{}\label{0formdual}
             \end{subfigure}
                          \hspace{29mm}

                    \begin{subfigure}[h]{0.49\textwidth}
                    \centering
  \includegraphics[width=0.9\textwidth]{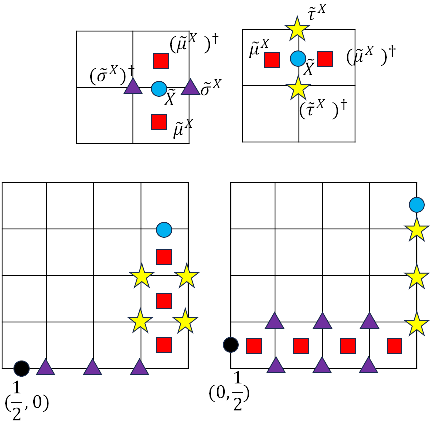}
         \caption{}\label{string3}
             \end{subfigure}
                               \begin{subfigure}[h]{0.49\textwidth}
                               \centering
  \includegraphics[width=1.1\textwidth]{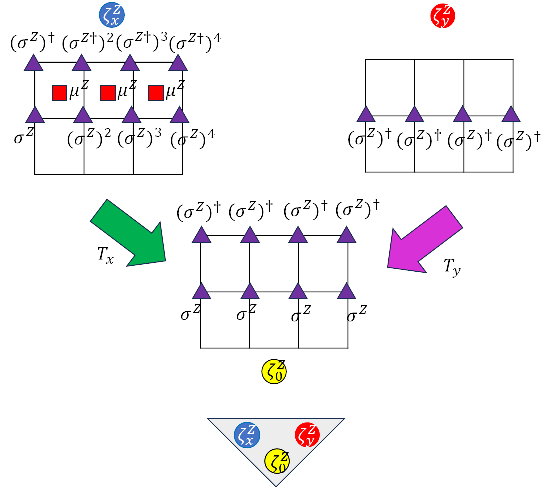}
         \caption{}\label{0form2}
             \end{subfigure}
                                         \begin{subfigure}[h]{0.49\textwidth}
                               \centering
  \includegraphics[width=0.9\textwidth]{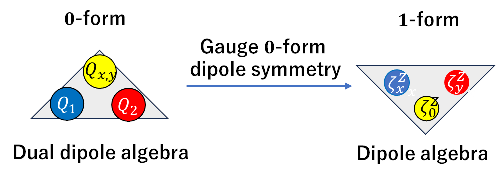}
         \caption{}\label{final}
             \end{subfigure}
 \end{center}
 \caption{(a)~Three types of terms defined in~\eqref{112} that respect the $0$-form dipole symmetry~\eqref{algebra3}. 
 (b)~$0$-form dipole symmetry, forming dual dipole algebra~\eqref{algebra3} which is schematically portrayed as an triangle in the bottom.
 (c)~Two configurations on the top correspond to the Gauss law defined in~\eqref{gauss3} whereas bottom two describe the non-local string of the gauge fields given in~\eqref{string4}. (d)~$1$-form operators introduced in~\eqref{lp4} constituting the $1$-form dipole algebra~\eqref{algebra4}.~(e) Duality found in this subsection: when gauging $0$-form dipole symmetry with dual dipole algebra, we have emergent $1$-form dipole symmetry with dipole algebra, where the hierarchical structure of the charges is inverted. 
 }
 \end{figure}
We can also think of a system with $0$-form dipole symmetry with the hierarchical structure of global and dipole charges is inverted, $0$-form dual dipole algebra and its noninvertible operators. 
Since the construction of the operator closely parallels the one in the previous subsections, we present how to do it succinctly. \par
We envisage a 2D lattice model where we accommodate two $\mathbb{Z}_N$ spin degrees of freedom on each link. The Hamiltonian is given by
\begin{equation}
\begin{split}
    H_{2D:dipole}^\prime&=-J_x\sum_{\br}\mathcal{K}^Z_{x,\br}-J_y\sum_{\br}\mathcal{K}^Z_{y,\br}-J_{xy}\sum_{\bp}\mathcal{K}^Z_{xy,\bp}-h_x\sum_{\bx}X_{\bx}-h_y\sum_{\by}X_{\by} \\
   & -\tl{J}_x\sum _{\br}\tl{\mathcal{K}}^Z_{x,\br}-\tl{J}_y\sum_{\br}\tl{\mathcal{K}}^Z_{y,\br}-J_{xy}\sum_{\bp}\tl{\mathcal{K}}^Z_{xy,\bp}-h_x\sum_{\bx}\tl{X}_{\bx}-h_y\sum_{\by}\tl{X}_{\by} +h.c.\label{114}
        \end{split}
\end{equation}
Here, we have introduced
\begin{eqnarray}
    \mathcal{K}^Z_{x,\br}\vcentcolon=Z^\dagger_{\br-\ex}Z_{\br+\ex},\quad \mathcal{K}^Z_{y,\br}\vcentcolon=Z^\dagger_{\br-\ey}Z_{\br+\ey},\quad
    \mathcal{K}^Z_{xy,\bp}\vcentcolon=Z_{\bp-\ex}Z^\dagger_{\bp+\ex}Z_{\bp+\ey}Z^\dagger_{\bp-\ey}\label{112}
\end{eqnarray}
and similarly for $ \tl{\mathcal{K}}^Z_{x,\br}$, $ \tl{\mathcal{K}}^Z_{y,\br}$, $ \tl{\mathcal{K}}^Z_{xy,\bp}$.
These spin coupling terms are demonstrated in Fig.~\ref{dualdipole}.
The model respects the following symmetries:
\begin{eqnarray}
    Q_{2D:1}=\prod_{\hx=1}^{L_x}\prod_{\hy=1}^{L_y}X_{\br+\ey},\quad Q_{2D:2}=\prod_{\hx=1}^{L_x}\prod_{\hy=1}^{L_y}X_{\br+\ex},\quad Q_{2D:x,y}=\prod_{\hx=1}^{L_x}\prod_{\hy=1}^{L_y}(X_{\br+\ey})^{\hx}(X_{\br+\ex})^{\hy},
    \label{algebra3}
\end{eqnarray}
and $\tl{Q}_{2D:1}$, $\tl{Q}_{2D:2}$, and $\tl{Q}_{2D:x,y}$ are analogously defined by replacing $X_{\bx}$ and $X_{\by}$ with $\tl{X}_{\bx}$ and $\tl{X}_{\by}$, respectively. The charges corresponding to such symmetries 
are depicted in Fig.~\ref{0formdual}.
Here, we have \textit{one} dipole charge and \textit{two} global charges, which is contrasted with the model in the previous subsection, where there are \textit{two} global charges and \textit{one} dipole charge. To see the distinction between the symmetries in the precious subsection and the one in the present case, we note that 
\begin{align}
    T_xQ_{2D:x,y}T_x^{-1}&=Q_{2D:x,y}Q_{2D:1}^\dagger,\nonumber \\  T_yQ_{2D:x,y}T_y^{-1}&=Q_{2D:x,y}Q_{2D:2}^\dagger, \label{dual00}\\
    T_xQ_{2D:i}T_x^{-1}&=  T_yQ_{2D:i}T_y^{-1}=  Q_{2D:i} \qquad(i=1,2), \nonumber
\end{align}
indicating that $   Q_{2D:1},   Q_{2D:2},   Q_{2D:x,y}$ constitute $0$-form dual dipole algebra, where the hierarchical structure of charges is inverted compared with the previous subsection~\eqref{spin2d2}~\eqref{algebra}. The analogous relation holds for another spin degree of freedom, i.e., spins with tilde.
The model further respects the $\mathbb{Z}_2$ symmetry exchanging two spin degrees of freedom. Defining the swap operator as
\begin{eqnarray}
    S_{\bx}=\frac{1}{N}\sum_{\alpha,\beta=1}^N\omega^{-\alpha\beta}\left[Z_{\bx}\tl{Z}_{\bx}^\dagger
     \right]^\alpha\times\left[X_{\bx}\tl{X}_{\bx}^{\dagger}\right]^\beta,\nonumber\\
    S_{\by}=\frac{1}{N}\sum_{\alpha,\beta=1}^N\omega^{-\alpha\beta}\left[Z_{\by}\tl{Z}_{\by}^\dagger
     \right]^\alpha\times\left[X_{\by}\tl{X}_{\by}^{\dagger}\right]^\beta,\label{you}
\end{eqnarray}
one verifies that the Hamiltonian~\eqref{114} commutes with $S$ with $S=\prod_{\hx=1}^{L_x}\prod_{\hy=1}^{L_y}  S_{\bx}S_{\by}$. The following discussion holds true as long as the Hamiltonian~\eqref{114} respects this $\mathbb{Z}_2$ symmetry and the one in~\eqref{algebra3}. For instance, 
one could add interaction terms to preserve these symmetries to the Hamiltonian. 
\par
Now we gauge this symmetry for the one of the spin degree of freedom, $\tl{Z}_{\bx}$, $\tl{Z}_{\by}$. To do so, we introduce two $\mathbb{Z}_N$ degrees of freedom on each node and one  $\mathbb{Z}_N$  on each plaquette, labeled by $\tl
{\tau}^{X/Z}_{\br}$, $\tl{\sigma}^{X/Z}_{\br}$, and $\tl{\mu}^{Z/X}_{\bp}$ corresponding to the gauge fields. The Gauss law reads
\begin{eqnarray}
    \tl{\sigma}^X_{\bx-\ex}   \tl{\sigma}^{X\dagger}_{\bx+\ex}\tl{\mu}^{X\dagger}_{\bx+\ey}\tl{\mu}^{X}_{\bx-\ey}\tl{X}_{\bx}=1,\quad \tl{\tau}^X_{\by+\ey}   \tl{\tau}^{X\dagger}_{\by-\ey}\tl{\mu}^{X\dagger}_{\by+\ex}\tl{\mu}^{X}_{\by-\ex}\tl{X}_{\by}=1,\label{gauss3}
\end{eqnarray}
which are portrayed in the top configurations in Fig.~\ref{string3}. Also, the spin coupling terms are minimally coupled with the gauge fields to commute with the Gauss law~\eqref{gauss3}, that is,
\begin{eqnarray}
     \tl{\mathcal{K}}^Z_{x,\br}\to\tl{\mathcal{K}}^Z_{x,\br}\tl{\sigma}^Z_{\br},\quad  \tl{\mathcal{K}}^Z_{y,\br}\to\tl{\mathcal{K}}^Z_{y,\br}\tl{\tau}^Z_{\br},\quad \tl{\mathcal{K}}^Z_{xy,\bp}\to\tl{\mathcal{K}}^Z_{xy,\bp}\tl{\mu}^Z_{\bp}
\end{eqnarray}
Defining new variables as
\begin{eqnarray*}
    \tau^X_{\br}\vcentcolon=\tl{\tau}^X_{\br}, \quad\sigma^X_{\br}\vcentcolon=\tl{\sigma}^X_{\br}, \quad\mu^X_{\bp}\vcentcolon=\tl{\mu}^X_{\bp}, \quad\tau^Z_{\br}\vcentcolon=\tl{\mathcal{K}}^Z_{y,\br}\tl{\tau}^Z_{\br},\quad \sigma^Z_{\br}\vcentcolon=\tl{\mathcal{K}}^Z_{x,\br}\tl{\sigma}^Z_{\br},\quad \mu^Z_{\bp}\vcentcolon=\tl{\mathcal{K}}^Z_{xy,\bp}\tl{\mu}^Z_{\bp}
\end{eqnarray*}
we have the following mapping via gauging the dipole symmetry:
\begin{eqnarray}
   \tl{ X}_{\bx}\Rightarrow G^X_{1,\bx},\quad   \tl{X}_{\by}\Rightarrow G^X_{2,\by},\quad \tl{\mathcal{K}}^Z_{x,\br}\Rightarrow \sigma^Z_{\br},\quad \tl{\mathcal{K}}^Z_{y,\br}\Rightarrow \tau^Z_{\br},\quad \tl{\mathcal{K}}^Z_{xy,\bp}\Rightarrow \mu^Z_{\bp}
\end{eqnarray}
where
\begin{eqnarray}
    G^X_{1,\bx}\vcentcolon=  {\sigma}^{X\dagger}_{\bx-\ex}   {\sigma}^{X}_{\bx+\ex}{\mu}^{X}_{\bx+\ey}{\mu}^{X\dagger}_{\bx-\ey},\quad   G^X_{2,\by}\vcentcolon={\tau}^{X\dagger}_{\by+\ey}   {\tau}^{X}_{\by-\ey}{\mu}^{X}_{\by+\ex}{\mu}^{X\dagger}_{\by-\ex}.
\end{eqnarray}
Note that the two operators $G^X_{1,\bx}$ and $G^X_{2,\by}$ are identical to the two flux operators $B_{\bx}$ and $B_{\by}$ given in~\eqref{fluxes} when replacing Pauli $Z$ operators with Pauli $X$ ones. We further add the following flux term to ensure that gauge theory is dynamically trivial, i.e., the flatness condition is imposed:
\begin{equation}
\begin{split}
&-g_B\sum_{\br} B_{\br} \vcentcolon=   \\
&-g_B\sum_{\br}{\tau^Z}_{\br-\exx}({\tau}^{Z\dagger}_{\br})^2{\tau^Z}_{\br+\exx}\times {\sigma}^Z_{\br-\eyy}({\sigma}^{Z\dagger}_{\br})^2 {\sigma}^Z_{\br+\eyy}\times ({\mu}^{Z}_{\bp-\ex+\ey})^\dagger{\mu}^{Z}_{\bp-\ex-\ey}({\mu}^Z_{\bp+\ex-\ey})^\dagger{\mu}^Z_{\bp+\ex+\ey}
 .\label{122}
 \end{split}
\end{equation}
This term is identical to $G_{\br}^X$ that was introduced in the previous subsection~\eqref{gsss} when replacing Pauli $Z$ operators with Pauli $X$ ones. \par
Overall, the gauged Hamiltonian reads
\begin{eqnarray}
    \widehat{H}_{2D:dipole}^\prime=\widehat{H}^\prime_0+\widehat{H}^\prime_{DTC}
\end{eqnarray}
with 
\begin{align}
\widehat{H}_{0}^\prime\vcentcolon=&-J_x\sum_{\br}\mathcal{K}^Z_{x,\br}-J_y\sum_{\br}\mathcal{K}^Z_{y,\br}-J_{xy}\sum_{\bp}\mathcal{K}^Z_{xy,\bp}-h_x\sum_{\bx}X_{\bx}-h_y\sum_{\by}X_{\by}
     -{J}_x\sum _{\br}\sigma_{\br}^Z-{J}_y\sum_{\br}\tau^Z_{\br}-J_{xy}\sum_{\bp}\mu^Z_{\bp}+h.c.\nonumber\\
 \widehat{H}^\prime_{DTC}\vcentcolon=& -h_x\sum_{\bx}G^X_{1,\bx}-h_y\sum_{\by}G^X_{2,\by}-g_B\sum_{\br}B_{\br}+h.c.
\end{align}
After gauging, we have $\widehat{H}^\prime_{DTC}$, which is
essentially the same Hamiltonian of the toric code with dipole symmetry~\eqref{89} up to exchanging Pauli $X$'s and Pauli $Z$'s. 
\par
Now we construct the operator. Similar to discussion presented in Sec.~\ref{423}, we replace local operators $\tl{Z}_{\bx}$ and $\tl{Z}_{\by}$ with 
the Wilson operators $W_{\bx}$ and $W_{\by}$
consisting of
 the string of the gauge fields~(see also two configurations in the bottom of Fig.~\ref{string3})~[recall that $\bx=(\hx+\frac{1}{2},\hy)$, and $\by=(\hx,\hy+\frac{1}{2})$]:
\begin{eqnarray}
    W_{\bx}&=&\prod_{\hx^\prime=1}^{\hx}\sigma^Z_{(\hx^\prime,0)} \times\prod_{\hy^\prime=0}^{\hy}\left[\tau^{Z\dagger}_{(\hx-1,\hy^\prime)}  \tau^{Z}_{(\hx+1,\hy^\prime)}\right]^{\hy^\prime}\mu^Z_{(\hx+\frac{1}{2},\hy^\prime+\frac{1}{2})}, \nonumber\\
    W_{\by}&=&\prod_{\hx^\prime=0}^{\hx}\left[\sigma^{Z\dagger}_{(\hx^\prime,0)} \sigma^{Z}_{(\hx^\prime,1)}\right]^{\hx^\prime}\mu^{Z\dagger}_{(\hx^\prime+\frac{1}{2},\frac{1}{2})} \times\prod_{\hy^\prime=1}^{\hy}\tau^Z_{(\hx,\hy^\prime)}. 
    \label{string4}
\end{eqnarray}
Substituting them into the swap operators~\eqref{you}, and introduce the product of such operators as
\begin{eqnarray}
    S=\left(\prod_{\bx\neq (\frac{1}{2},0)}\prod_{\by\neq (0,\frac{1}{2})}S_{\bx}S_{\by}\right)\times S_{(\frac{1}{2},0)}S_{(0,\frac{1}{2})}.
\end{eqnarray}
We check how it acts on the spins to see whether it behaves the mapping corresponding gauging. In some cases, such as the one when $S$ acts on spin around the boundary, it picks up non-local charges.
Omitting the details of this discussion as it closely parallels the ones in the previous argument (Sec.~\ref{423}), it turns out that to have the proper operator, one needs to multiply $S$ with 
some of the operators, that is,
\begin{eqnarray}
    D=\frac{1}{N^3}S\times C
\end{eqnarray}
with $C=C^\prime_0C^\prime_1$ and 
\begin{eqnarray}
C^\prime_0\vcentcolon&=&\sum_{\alpha,\beta,\gamma=1}^N Q_{2D:1}^\alpha Q_{2D:2}^\beta Q_{2D:x,y}^\gamma,\nonumber\\
C^\prime_1\vcentcolon&=&\frac{1}{N^{L_xL_y}}\sum_{\gamma}\eta(\gamma).\label{ce}
\end{eqnarray}
Here, three charges $Q_{2D:1}$, $Q_{2D:2}$, $Q_{2D:x,y}$
are given in~\eqref{algebra3}, which form dual dipole algebra and
$\sum_{\gamma}$ stands for summing all kinds of loops of the gauge fields $\tau^Z_{\br}$, $\sigma^Z_{\br}$, $\mu^Z_{\bp}$, including noncontractible ones. When taking the limit of $g_B\to \infty$, $C^\prime_1$ becomes (see also Appendix.~\ref{app:1})
\begin{eqnarray}
    C^\prime_1=\frac{1}{N^3}\sum_{p,q,r=1}^N\sum_{s,t,u=1}^N(\zeta_{0}^Z)^p (\zeta_{x}^Z)^q(\zeta_{y}^Z)^r
   ( \eta_{0}^Z)^s (\eta_{x}^Z)^t(\eta_{y}^Z)^u.\label{cond3}
\end{eqnarray}
Here, the first three terms denote the non-contractible loops of the gauge fields in the $x$-direction whereas the last three do in the $y$-direction, namely, 
\begin{eqnarray}
\zeta_{0}^Z\vcentcolon=\prod_{\hx=1}^{L_x}\sigma^{Z\dagger}_{(\hx,0)}\sigma^{Z}_{(\hx,1)},\quad \zeta_{x}^Z\vcentcolon=\prod_{\hx=1}^{L_x}(\sigma^{Z\dagger}_{(\hx,0)})^{\hx}(\sigma^{Z}_{(\hx,1)})^{\hx}\mu^{Z\dagger}_{(\hx+\frac{1}{2},\frac{1}{2})},\quad \zeta_{y}^Z\vcentcolon=\prod_{\hx=1}^{L_x}(\sigma_{(\hx,0)}^Z)^\dagger,\label{lp4}
\end{eqnarray}
which are portrayed in Fig.~\ref{0form2}. When $g_B\to\infty$,
these loops are subject to the following relation:
\begin{align}
     T_x\zeta_{x}^ZT_x^{-1}&=\zeta_{x}^Z\zeta_{0}^{Z\dagger},&&\quad&&
    T_y\zeta_{y}^ZT_y^{-1}=\zeta_{y}^Z\zeta_{0}^{Z\dagger}, \nonumber\\
   T_x\zeta_{y}^ZT_x^{-1}&=\zeta_{y}^Z,\quad&& T_y\zeta_{x}^ZT_y^{-1}=\zeta_{x}^Z,\quad&&   T_x\zeta_{0}^ZT_x^{-1}= T_y\zeta_{0}^ZT_y^{-1}=\zeta_{0}^Z,
    \label{algebra4}
\end{align}
implying that the loops constitute $1$-form dipole algebra with the hierarchical structure is inverted compared with the one in the original $0$-form dipole symmetry~(Fig.~\ref{final})
. The last two relations in~\eqref{algebra4} are obtained by the fact that the operators are topological, i.e., independent operators depend solely on the homology class.
\footnote{
To see this more explicitly, we derive 
the fourth relation in~\eqref{algebra4}. We start with
\begin{eqnarray*}
     T_y\zeta_{x}^ZT_y^{-1}=\prod_{\hx=1}^{L_x}(\sigma^{Z\dagger}_{(\hx,1)})^{\hx}(\sigma^{Z}_{(\hx,2)})^{\hx}\mu^{Z\dagger}_{(\hx+\frac{1}{2},\frac{3}{2})}.
\end{eqnarray*}
The right hand side is equivalent to $\prod_{\hx=1}^{L_x}(\sigma^{Z\dagger}_{(\hx,0)})^{\hx}(\sigma^{Z}_{(\hx,1)})^{\hx}\mu^{Z\dagger}_{(\hx+\frac{1}{2},\frac{1}{2})}=\zeta^Z_x$
, obtained by the fact that we focus on the ground state satisfying $B_{\br}\ket{\Omega}=\ket{\Omega}$ with $B_{\br}$ being given by~\eqref{122}. The last relation in~\eqref{algebra4} can be analogously derived.}

Likewise, the last three loops that enter in~\eqref{cond3} are given by
\begin{eqnarray}
    \eta^Z_{0}\vcentcolon=\prod_{\hy=1}^{L_y}\tau^{Z\dagger}_{(0,\hy)}\tau^Z_{(1,\hy)},\quad \eta^Z_{x}\vcentcolon=\prod_{\hy=1}^{L_y}\tau^{Z\dagger}_{(0,\hy)},\quad \eta^Z_{y}\vcentcolon=\prod_{\hy=1}^{L_y}(\tau^{Z\dagger}_{(0,\hy)})^{\hy}(\tau^Z_{(1,\hy)})^{\hy}\mu^{Z\dagger}_{(\frac{1}{2},\hy)}
\end{eqnarray}
which are subject to the relation obtained by replacing $\zeta^Z_{0}$, $\zeta^Z_{x}$, and $\zeta^Z_{y}$ with $\eta^Z_{0}$, $\eta^Z_{x}$, and $\eta^Z_{y}$, respectively in~\eqref{algebra4}.
\par
The operator $D$ is the desired swap operator whose action on local spins and spin couplings reads
\begin{eqnarray}
    DX_{\bx}=G^X_{1,\bx} D,\quad   DX_{\by}=G^X_{2,\by} D,\quad D\mathcal{K}^Z_{x,\br}=\sigma^Z_{\br} D,\quad D\mathcal{K}^Z_{y,\br}=\tau^Z_{\br}D,\quad D\mathcal{K}^Z_{xy,\bp}=\mu^Z_{\bp}D\nonumber\\
    DG^X_{\bx}=X_{\bx}D,\quad  DG^X_{\by}=X_{\by}D,\quad D\sigma^Z_{\br}=\mathcal{K}^Z_{x,\br}D,\quad D\tau^Z_{\br}=\mathcal{K}^Z_{y,\br}D,\quad D\mu^Z_{\bp}=\mathcal{K}^Z_{xy,\bp}D.
\end{eqnarray}
Furthermore, this operator is noninvertible; one obtains the following fusion rules:
\begin{eqnarray}
    D\times D=C,\quad C\times C=N^6C,\quad C\times D=D\times C=N^6D.\label{fusiondip3}
\end{eqnarray}
Especially, when $g_B\to \infty$, $C$ becomes
\begin{eqnarray}
    C=\left(\sum_{\alpha,\beta,\gamma=1}^N Q_{2D:1}^\alpha Q_{2D:2}^\beta Q_{2D:x,y}^\gamma\right)\times\left(\frac{1}{N^3}\sum_{p,q,r=1}^N\sum_{s,t,u=1}^N(\zeta_{0}^Z)^p (\zeta_{x}^Z)^q(\zeta_{y}^Z)^r
   ( \eta_{0}^Z)^s (\eta_{x}^Z)^t(\eta_{y}^Z)^u\right),\label{fusiondip4}
\end{eqnarray}
indicating that $C$ involves $0$-form dipole symmetry with dual dipole algebra and $1$-form dipole symmetry with dipole algebra. 
Although the fusion rules~\eqref{fusiondip3} look resemble the ones in~\eqref{fusiondip}, the content of $C$ is different: in the present case, $C$ contains $0$-form dipole symmetry with dual dipole algebra and $1$-form one with dipole algebra
~(compare also Fig.~\ref{dl} and Fig.~\ref{final}).
\section{Outlook}
\label{section5}
Stimulated by recent development of the two types of exotic symmetries---noninvertible and spatially modulated symmetries, we explore the interplay between them via lattice gauge theories. Introducing two copies of spin models on a lattice defined in one, two, and three dimensions, 
we demonstrate a systematic approach to construct noninvertible operators by
gauging subsystem or dipole symmetries, which are typical examples of the spatially modulated symmetries that have emerged in fractonic topological phases.
Such constructions also help better understand noninvertible operators in higher dimensional lattice models, which are active areas of research in communities of high energy and condensed matter physics. 
\par
In the case of subsystem symmetries, we have constructed 
noninvertible operators whose fusion rules give rise to subsystem charges. 
In two dimensions, such fusion rules correspond to the subsystem analog of the category theory forming $\text{Rep}(D_8)$. In three dimensions, fusion rules of the operators involve $0$-form subsystem charges and higher-form operators, associated with lineon excitations found in the X-cube model. The operator is the subsystem analog of 
$2$-Rep$\left((\mathbb{Z}_2^{(1)}\times\mathbb{Z}_2^{(1)})\rtimes\mathbb{Z}_2^{(0)}\right)$ discussed in the fusion $2$-category theory~\cite{bhardwaj2023non,Choi:2024rjm}.
\par
In the case of dipole symmetries, especially in two dimensions, a new degree of freedom has been taken into account --- dipole algebra, comprised of global and dipole charges. We show that gauging $0$-form
dipole symmetry with a given dipole algebra in two dimensions gives rise to $1$-form \textit{dual dipole algebra} where the hierarchical structure of the algebra is inverted. 
Such a finding is the generalization of the known fact that gauging a global symmetry generates an emergent quantum symmetry, to a dipole symmetry.
This new type of duality is also manifest in the fusion rules of the noninvertible operators. Indeed, the fusion rule involves $0$-form dipole symmetry forming a dipole algebra given a hierarchy and $1$-form dipole symmetry with the hierarchical stricture is inverted (Figs.~\ref{dl} and~\ref{final}).
Our results would comply with recent interests in construction of a unified framework to incorporate various kinds of symmetries, including noninvertible and spatially modulated ones.\par

One can extend our analysis in many different ways, which we leave for future studies. 
First, it is interesting to explore gauging more generic group symmetries, in addition to $\mathbb{Z}_N$ symmetries. In particular, the non-Abelian dipole symmetry has not been discussed yet. \par
Second, 
one could study higher multipole symmetries in various dimensions. 
Especially, in dimensions higher than two, such as three dimensions, it is known that gauging $0$-form dipole symmetry with dipole algebra and the one with dual dipole algebra lead to different topological orders compared with the two dimensional analog~\cite{anomaly_2024}. Exploring richer structures of (dual) multipole algebras thereof in various dimension provides further insight into the spatially modulated symmetries. \par
Third, systematic investigations on the effects of various boundary conditions should be performed. Specifically, when $N \mid L$ (meaning, when the system size $L$ is not divisible by $N$), we can impose twisted boundary conditions for various symmetries, which affects the ground-state degeneracy of the system; when $N \nmid L$, duality twisted boundary conditions~\cite{Schutz1992twistbc} are required, which can lead to exotic topological properties and new insight at critical points~\cite{YuChenRoyTeo2017defect}. \par
Fourth, it could be an interesting and important problem on its own right to explore phase diagram of the spin models after gauging the spatially modulated symmetries. Especially, after gauging, one could add matter terms coupled with the gauge fields in addition to other terms such as the ones constitute topological model with the modulated symmetries, reminiscent of the Fradkin-Shenker model~\cite{fradkin_shenker}. Identifying possible phases in the model and investigating the dualities of the modulated symmetries 
in these phases allow one to gain more fruitful insights on the modulated symmetries. \par
Fifth, it would be intriguing to explore noninvertible duality defects that are generated by gauging the entire symmetry group; by contrast, our noninvertible operators are obtained by gauging a proper subgroup (modulated symmetries for a single copy of spin systems instead of both copies) of the entire symmetry group.
Generally, the noninvertible duality defect in $d$ spatial dimension is obtained when the theory is invariant under the gauging $p$-form symmetry.
Especially, in the case of the dipole symmetry, the conditions to have such a defect are (i) $d=(p+1)/2$ (ii)~The dipole symmetry has the same dipole algebra before and after gauging. It would be also interesting to address whether one can extend our analysis beyond square lattices and examine how dipole symmetry and its algebra is generalized by incorporating other types of spatial symmetries such as rotations and refections.

Last but not least, establishing category theory corresponding to the fusion rules of the noninvertible operators discovered in this paper will fascinate both physicists and mathematicians. 
\section*{Acknowledgement}
We would like to thank 
T.~Ando,~J. H. Han, M. Honda,
R. Kalloor, 
A. Nayak,
T. Nakanishi,~T. Ohishi,
T. Saito,~S. Shimamori, S.~Shimomura,  
K. Shiozaki, T.~Takama, and
P. Tanay for helpful discussion. 
We also thank W. Cao and A. Tiwari for useful comments on the draft and suggesting relevant works that we have missed.
This work is in part supported by KAKENHI-PROJECT-23H01097, the Koshland Fellowship at the Weizmann Institute of Science, JST CREST Grant Number JPMJCR24I3. \par


\appendix

\section{Condensation defects}\label{app:1}
In this Appendix, we give detailed explanations to derive the relations~\eqref{cond},~\eqref{cond2}, and~\eqref{cond3}.

\subsection{Subsystem symmetry in three dimensions}
We first give derivation of~\eqref{cond}.
Recall that an operator $C_{\eta}$~\eqref{eeta} is introduced when constructing operators obtained by gauging subsystem symmetries in three dimensions, which has the form
\begin{eqnarray*}
     C_{\eta}=\frac{1}{2^{2L_xL_yL_z}}\sum_{\gamma}\eta(\gamma),
\end{eqnarray*}
where the sum is taken for all kinds of loops of the gauge fields, including noncontractible ones. 
We will separate $C_\eta$ into the terms comprised of the contractible loops and the ones of the noncontractible. To this end, 
we first note that there are subextensive number of redundancies regarding the flux operators,~$B_{\bc,x}$,~$B_{\bc,y}$ given in~\eqref{flux}. Indeed, 
we have the following $L_x+L_y+L_z-1$ constraints:
\begin{eqnarray}
   \prod_{\hy^\prime=1}^{L_y}  \prod_{\hz^\prime=1}^{L_z}B_{(\hx,\hy^\prime,\hz^\prime),x}=1~(1\leq \hx\leq L_x),&\quad    \prod_{\hx^\prime=1}^{L_x}  \prod_{\hz^\prime=1}^{L_z}B_{(\hx^\prime,\hy,\hz^\prime),y}=1~(1\leq \hy\leq L_y),\nonumber\\
   \prod_{\hx^\prime=1}^{L_x}  \prod_{\hy^\prime=1}^{L_y}B_{(\hx^\prime,\hy\prime,\hz),x}B_{(\hx^\prime,\hy\prime,\hz),y}=1&~(1\leq \hz\leq L_z-1).\label{ff}
\end{eqnarray}
Now we think of the following product:
\begin{eqnarray}
    \prod_{\bc}\frac{1}{2^2}(1+B_{\bc,x})(1+B_{\bc,y}),
\end{eqnarray}
which can be rewritten as 
\begin{eqnarray}
     \prod_{\bc}\frac{1}{2^2}(1+B_{\bc,x})(1+B_{\bc,y})=\prod_{\bc\neq \{\bc_i\}}\frac{1}{2^2}(1+B_{\bc,x})(1+B_{\bc,y}),
\end{eqnarray}
where $B_{\bc^\prime,x}$ and $B_{\bc^\prime,y}$ with
$\bc^\prime\in\{\bc_i\}$ describes distinct $L_x+L_y+L_z-1(\vcentcolon=K)$ flux operators each of which enters the $K$ different constraints given in~\eqref{ff}.
Further, one finds
\begin{eqnarray}
    \prod_{\bc\neq \{\bc_i\}}\frac{1}{2^2}(1+B_{\bc,x})(1+B_{\bc,y})=\frac{1}{2^{2L_xL_yL_z-K}}\sum_{\gamma_0}\eta(\gamma_0),\label{ap2}
\end{eqnarray}
where the sum on the right hand side is taken for contractible loops. Since we have just identified contrition of the contractible loops in  $C_{\eta}$, it follows that 
\begin{align}
    C_\eta&=\frac{1}{2^K}\left[\frac{1}{2}\prod_{\hz=1}^{L_z}\left(1+\eta^{zx}_x(\hz)\right)\left(1+\eta^{yz}_y(\hz)\right)\right]\times \left[\frac{1}{2}\prod_{\hx=1}^{L_x}\left(1+\eta^{zx}_z(\hx)\right)\left(1+\eta^{xy}_y(\hx)\right)\right]\nonumber\\
& \qquad \times \left[\frac{1}{2}\prod_{\hy=1}^{L_y}\left(1+\eta^{xy}_x(\hy)\right)\left(1+\eta^{yz}_z(\hy)\right)\right]\times \frac{1}{2^{2L_xL_yL_z-K}}\sum_{\gamma_0}\eta(\gamma_0),
\end{align}
where $\eta^{ab}_a(\hat{c})$ defined in~\eqref{ab} denotes the noncontractible loops of the lineon excitations. \par
When $g_x,g_y,g_z\to\infty$, the conditions $B_{\bc,x}=1$ and $B_{\bc,y}=1$ are strictly imposed. From~\eqref{ap2}, it follows that 
$\frac{1}{2^{2L_xL_yL_z-K}}\sum_{\gamma_0}\eta(\gamma_0)=1$, leading to
\begin{align*}
     C_\eta&=\frac{1}{2^K}\left[\frac{1}{2}\prod_{\hz=1}^{L_z}\left(1+\eta^{zx}_x(\hz)\right)\left(1+\eta^{yz}_y(\hz)\right)\right]\times \left[\frac{1}{2}\prod_{\hx=1}^{L_x}\left(1+\eta^{zx}_z(\hx)\right)\left(1+\eta^{xy}_y(\hx)\right)\right]\nonumber\\
& \qquad \times \left[\frac{1}{2}\prod_{\hy=1}^{L_y}\left(1+\eta^{xy}_x(\hy)\right)\left(1+\eta^{yz}_z(\hy)\right)\right],
\end{align*}
which is nothing but~\eqref{cond}.
\subsection{Dipole symmetry in two dimensions}
We turn to the derivation of~\eqref{cond2}. When constructing operators via gauging dipole symmetry in two dimensions, an operator $C_1$~\eqref{c1} is introduced, which has the form
\begin{eqnarray*}
      C_{1}=\frac{1}{N^{2L_xL_y}}\sum_{\gamma}\eta(\gamma).
\end{eqnarray*}
We will decompose it into contribution from contractible loops and the one from noncontractible. To this end, we note that there are three constraints regarding the flux operators, $B_{\bx}$, $B_{\by}$~\eqref{fluxes}, reading
\begin{eqnarray}
    \prod_{\hx=1}^{L_x} \prod_{\hy=1}^{L_y}B_{\bx}=1,\quad   \prod_{\hx=1}^{L_x} \prod_{\hy=1}^{L_y}B_{\by}=1,\quad  \prod_{\hx=1}^{L_x} \prod_{\hy=1}^{L_y}(B_{\bx})^{\hy}(B_{\by})^{\hx}=1\label{shs}
\end{eqnarray}
We think of the following product:
\begin{eqnarray}
    \prod_{\bx}\prod_{\by}\frac{1}{N^2}\left(\sum_{\alpha=1}^NB_{\bx}^\alpha\right)\times \left(\sum_{\beta=1}^NB_{\by}^\beta\right),
\end{eqnarray}
which can be further written as 
\begin{eqnarray}
         \prod_{\bx}\prod_{\by}\frac{1}{N^2}\left(\sum_{\alpha=1}^NB_{\bx}^\alpha\right)\times \left(\sum_{\beta=1}^NB_{\by}^\beta\right)= \prod_{\bx\neq \mathbf{l}_{x0},\mathbf{l}_{x1}}\prod_{\by\neq \mathbf{l}_{y0}}\frac{1}{N^2}\left(\sum_{\alpha=1}^NB_{\bx}^\alpha\right)\times \left(\sum_{\beta=1}^NB_{\by}^\beta\right),\label{143}
\end{eqnarray}
where $B_{\mathbf{l}_{x0}}$, $B_{\mathbf{l}_{x1}}$, and $B_{\mathbf{l}_{y0}}$ represent three different flux operators each of which enters the three different constraints presented in~\eqref{shs}. Furthermore,~\eqref{143} can be transformed into
\begin{eqnarray}
    \prod_{\bx\neq \mathbf{l}_{x0},\mathbf{l}_{x1}}\prod_{\by\neq \mathbf{l}_{y0}}\frac{1}{N^2}\left(\sum_{\alpha=1}^NB_{\bx}^\alpha\right)\times \left(\sum_{\beta=1}^NB_{\by}^\beta\right)=\frac{1}{N^{2L_xL_y-3}}\sum_{\gamma_0}\eta(\gamma_0),\label{qqq0}
\end{eqnarray}
where on the right hand side we sum over contractible loops of the gauge fields. Since we have identified the contribution from the contractible loops in~\eqref{c1}, the remaining degree of freedom is the noncontractible loops. Hence, we have
\begin{eqnarray}
    C_1=\left(\frac{1}{N^{2L_xL_y-3}}\sum_{\gamma_0}\eta(\gamma_0)\right)\times\left(\frac{1}{N^3}\sum_{p,q,r=1}^N\sum_{s,t,u=1}^N(\xi_{x,y}^Z)^p (\xi_{1}^Z)^q(\xi_{2}^Z)^r
   ( \eta_{x,y}^Z)^s (\eta_{1}^Z)^t(\eta_{2}^Z)^u\right).
\end{eqnarray}
Here, six operators inside the second braket are given in~\eqref{loops}~\eqref{loopy}. When $g_{B_x},g_{B_y}\to\infty$, the condition $B_{\bx}=B_{\by}=1$ is strictly imposed, leading to $\frac{1}{N^{2L_xL_y-3}}\sum_{\gamma_0}\eta(\gamma_0)=1$ according to~\eqref{qqq0}. Therefore we obtain~\eqref{cond2}:
\begin{eqnarray*}
      C_1=\frac{1}{N^3}\sum_{p,q,r=1}^N\sum_{s,t,u=1}^N(\xi_{x,y}^Z)^p (\xi_{1}^Z)^q(\xi_{2}^Z)^r
   ( \eta_{x,y}^Z)^s (\eta_{1}^Z)^t(\eta_{2}^Z)^u.
\end{eqnarray*}
\par
One can also derive the relation~\eqref{cond3} discussed in the case of the two dimensional spin model with dipole symmetry with dual dipole algebra
by the following the similar logic presented in this subsection. One can do so by noticing the fact that there are three constraints regarding the flux operator~$B_{\br}$ defined in~\eqref{122}:
\begin{eqnarray}
    \prod_{\hx=1}^{L_x} \prod_{\hy=1}^{L_y}B_{\br}=1,\quad   \prod_{\hx=1}^{L_x} \prod_{\hy=1}^{L_y}(B_{\br})^{\hx}=1,  \quad \prod_{\hx=1}^{L_x} \prod_{\hy=1}^{L_y}(B_{\br})^{\hy}=1.
\end{eqnarray}

\bibliography{main}
\bibliographystyle{utphys}
\end{document}